%% file: main.tex
\documentclass[10pt,conference]{IEEEtran}
\usepackage{cite}
\usepackage{rotating}
\usepackage{amsmath,amssymb,amsfonts}
\usepackage{algorithmic}
\usepackage{booktabs}
\usepackage[table,xcdraw]{xcolor}
\usepackage{graphicx}
\usepackage{subcaption}
\usepackage{textcomp}
\usepackage[most]{tcolorbox}
\usepackage{xcolor}
\usepackage{textcomp}
\usepackage{multirow}
\usepackage{lscape}
\usepackage{longtable}
\usepackage{tabu}
\usepackage{url}
\usepackage{fontawesome}
\usepackage{adjustbox}

\usepackage{breakurl}
\usepackage[breaklinks]{hyperref}
\def\BibTeX{{\rm B\kern-.05em{\sc i\kern-.025em b}\kern-.08em
    T\kern-.1667em\lower.7ex\hbox{E}\kern-.125emX}}

\begin{document}

\title{
Towards Concrete and Connected AI Risk Assessment (C$^2$AIRA): A Systematic Mapping Study
}


\author{
\IEEEauthorblockN{Boming Xia\IEEEauthorrefmark{1}\IEEEauthorrefmark{2}, Qinghua Lu\IEEEauthorrefmark{1}\IEEEauthorrefmark{2}, Harsha Perera\IEEEauthorrefmark{1}, Liming Zhu\IEEEauthorrefmark{1}\IEEEauthorrefmark{2}, Zhenchang Xing\IEEEauthorrefmark{1}, Yue Liu\IEEEauthorrefmark{1}\IEEEauthorrefmark{2}, Jon Whittle\IEEEauthorrefmark{1}}
\IEEEauthorblockA{\IEEEauthorrefmark{1}\textit{CSIRO's Data61, Sydney, Australia} \\
\IEEEauthorrefmark{2}\textit{University of New South Wales, Sydney, Australia}
}
}


\maketitle

\begin{abstract}
The rapid development of artificial intelligence (AI) has led to increasing concerns about the capability of AI systems to make decisions and behave responsibly.
Responsible AI (RAI) refers to the development and use of AI systems that benefit humans, society, and the environment while minimising the risk of negative consequences. To ensure responsible AI, the risks associated with AI systems’ development and use must be identified, assessed and mitigated.
Various AI risk assessment frameworks have been released recently by governments, organisations, and companies.
However, it can be challenging for AI stakeholders to have a clear picture of the available frameworks and determine the most suitable ones for a specific context. Additionally, there is a need to identify areas that require further research or development of new frameworks, as well as updating and maintaining existing ones.
To fill the gap, we present a mapping study of 16 existing AI risk assessment frameworks from the industry, governments, and non-government organizations (NGOs). We identify key characteristics of each framework and analyse them in terms of RAI principles, stakeholders, system lifecycle stages, geographical locations, targeted domains, and assessment methods. Our study provides a comprehensive analysis of the current state of the frameworks and highlights areas of convergence and divergence among them.
We also identify the deficiencies in existing frameworks and outlines the essential characteristics of a concrete and connected framework AI risk assessment (C$^2$AIRA) framework. Our findings and insights can help relevant stakeholders choose suitable AI risk assessment frameworks and guide the design of future frameworks towards concreteness and connectedness.

\end{abstract}


\section{Introduction}
\label{Sec:Intro}
\input{1Introduction.tex}

\section{Methodology}
\label{Sec:Methodology}
\input{2Methodology.tex}

\section{Terminology}
\label{Sec:Term}
In the section, we define and explain some key terms in this study for clarity.

\textbf{Responsible AI}. The term ``responsible AI" is often used interchangeably with other related terms such as ``ethical AI", ``trustworthy AI", ``AI for Good", ``values-driven AI", and, more broadly, ``digital humanism". However, despite their nuances, these terms all share a common goal: to promote the development, deployment, and use of AI systems that have a positive impact on individuals, groups, and society while minimizing associated risks.

\textbf{AI system}. In this study, an AI system holistically refers to systems or machines that contain AI components to perform tasks that would normally require human intelligence, such as natural language processing and image recognition.

\textbf{AI risk}. AI risk encompasses the potential for both intended and unintended harm or adverse consequences arising from any AI-related activities, such as research, development, deployment, operation, maintenance, and procurement of AI systems. We further categorize AI risks into hazard, exposure, vulnerability, and mitigation risks in Section \ref{Sec:RQ2}.

\textbf{RAI principles' relation to AI risks}. The RAI principles can be considered as quality metrics for operationalizing RAI.
\begin{itemize}
    \item HSE wellbeing: This principle aims to mitigate risks related to negative impacts on these areas.
    \item Human-centered values: This principle aims to mitigate risks related to violations of human rights and dignity.
    \item Fairness: This principle aims to prevent bias and discrimination in AI systems, thereby mitigating risks related to unfair treatment of individuals or groups.
    \item Privacy protection and security: This principle helps mitigate risks related to data breaches and misuse.
    \item Reliability and safety: This principle ensures that AI systems are reliable and safe to use, thereby mitigating risks related to unreasonable or unintended system failures or accidents caused by AI.
    \item Transparency and explainability: This principle helps mitigate risks related to lack of understanding or trust in AI systems, by ensuring that AI systems are transparent in their operations and can be explained to relevant stakeholders if needed.
    \item Contestability: This principle promotes accountability and helps mitigate risks related to lack of recourse for individuals (especially the vulnerable ones) affected by AI decisions, by ensuring that individuals have the right to challenge decisions made by AI systems and seek redress for negative impacts caused by AI.
    \item Accountability: This principle helps mitigate risks related to lack of accountability for negative impacts caused by AI, by ensuring that there is clear responsibility identification for different AI system lifecycle stages and the actions of AI systems.
\end{itemize}

\section{Research results}
\label{Sec:results}
\input{3Results}

\section{Discussion}
\label{Sec:discussion}
\input{4Discussion}

\section{Related work}
\label{Sec:relatedwork}

\input{5RelatedWork}

\section{Conclusion and future work}
\label{Sec:Conclusion}
This paper presents a systematic mapping study that aims to evaluate the effectiveness and limitations of existing AI risk assessment frameworks. Through the mapping study, we provide a comprehensive overview of the key characteristics (i.e., qualities, elements, and processes) of a C$^2$AIRA framework, which includes well-defined RAI principles, RAI stakeholders, AI system lifecycle stages, applicable sectors and regions, risk factors, and reusable mitigations. Additionally, we offer valuable insights to facilitate the development of C$^2$AIRA frameworks with assessment and mitigation measures in an interconnected and layered manner.

As a part of future work, we are developing a question bank with risk assessment questions labeled with respect to different aspects (e.g., RAI principles, RAI stakeholders, lifecycle stages) to provide a better foundation for the development of C$^2$AIRA frameworks.


\Urlmuskip=0mu plus 1mu\relax
\bibliographystyle{IEEEtran}
\bibliography{main}

\end{document}

%% file: 1Introduction.tex
The adoption of artificial intelligence (AI) in various application domains has led to numerous advantages, such as improved efficiency and reduced cost in manufacturing. However, the risks associated with AI systems have also attracted significant attention from both industry and academia \cite{10.1145/3522664.3528607, sambasivan2021everyone, gao2019towards}.
For example, an AI system may make biased decisions that lead to unintended discrimination\cite{ferrer2021bias, nasim2022artificial, nagbol2021designing}. Also, the AI system's dataset may contain sensitive information, risking violation of laws such as \href{https://gdprinfo.eu/}{EU General Data Protection Regulation (GDPR)}\footnote{\url{https://gdprinfo.eu/}} and \href{https://artificialintelligenceact.eu/}{EU AI Act (proposed)}\footnote{\url{https://artificialintelligenceact.eu/}}.
The \href{https://incidentdatabase.ai/}{AI incident database}\footnote{\url{https://incidentdatabase.ai/}} has collected over 2200 (as of January 2023) reported real-world incidents caused by AI systems.

Responsible AI (RAI) is the practice of developing, deploying, and using AI systems in a way that benefits individuals, groups, and society at large, while minimizing the risk of negative consequences \cite{RAIDef}. A number of RAI principle frameworks that AI systems and stakeholders should adhere to have been released recently, such as \href{https://www.industry.gov.au/publications/australias-artificial-intelligence-ethics-framework/australias-ai-ethics-principles}{Australia's AI Ethics Principles}\footnote{\url{https://www.industry.gov.au/publications/australias-artificial-intelligence-ethics-framework/australias-ai-ethics-principles}} and \href{https://ec.europa.eu/digital-single-market/en/news/ethics-guidelines-trustworthy-ai}{European Commission's Ethics guidelines for trustworthy AI}\footnote{\url{https://ec.europa.eu/digital-single-market/en/news/ethics-guidelines-trustworthy-ai}}, which AI systems and stakeholders should adhere to. To implement RAI based on these principles, many organizations have developed principle-driven AI risk assessment frameworks (e.g., US NIST AI risk management framework\cite{NIST_AIRMF}, EU Assessment List for Trustworthy AI\cite{EUALTAI}). These frameworks are designed to help organizations and individuals systematically identify and mitigate potential risks associated with AI systems. 
Despite the availability of these AI risk assessment frameworks, it is important for AI system stakeholders to have a comprehensive understanding of the existing frameworks to choose the most appropriate one for their context. Moreover, it is unclear how effective these frameworks are at assessing and mitigating AI risks.
While there are studies or online repositories (e.g., \cite{EYreport, AIguidelines, AIinventory}) that include RAI resources such as guidelines, frameworks, policies, and standards, these resources primarily serve as collections or overviews without focusing on more concrete AI risk assessment frameworks.

To bridge the gaps, we have performed a systematic mapping study on the existing AI risk assessment frameworks.
The main objectives of this study are: 1) to provide a summary of the current available higher-quality AI risk frameworks to which researchers and practitioners can refer; 2) to investigate the capabilities and limitations of the AI risk assessment frameworks; and 3) to provide insights for future research and development on concrete and connected AI risk assessment (C$^2$AIRA) frameworks.

The main contributions of this study are:

\begin{itemize}
    \item We present a comprehensive qualitative and quantitative analysis and synthesis of 16 state-of-practice AI risk assessment frameworks selected from the grey literature. 
    \item We provide empirical findings and insights on 
    the capabilities and limitations of the existing frameworks and highlight the essentials for developing C$^2$AIRA frameworks.
\end{itemize}

This paper is organized as follows: Section~\ref{Sec:Methodology} outlines the methodology and research questions (RQs). Section~\ref{Sec:Term} clarifies key terms and the relationships between AI risks and RAI principles. The results and findings for each RQ are presented in Section~\ref{Sec:results}. Section~\ref{Sec:discussion} discusses the concepts of ``concreteness" and ``connectedness" for C$^2$AIRA, as well as potential threats to validity. Section~\ref{Sec:relatedwork} provides an overview of related work, while Section~\ref{Sec:Conclusion} concludes the paper with a summary and future work.

%% file: 2Methodology.tex
We perform the systematic mapping study following the guidelines on conducting systematic (multivocal) literature reviews \cite{Garousi_2019, kitchenham2009systematic} and mapping studies \cite{Petersen_2015} in software engineering. The overall methodology is presented in Figure \ref{fig:methodology}. To investigate the capabilities and limitations of the existing AI risk assessment frameworks, we derived the following RQs in terms of the characteristics/features and the assessment processes:

\begin{figure}[]
\centering
\includegraphics[width=\columnwidth]{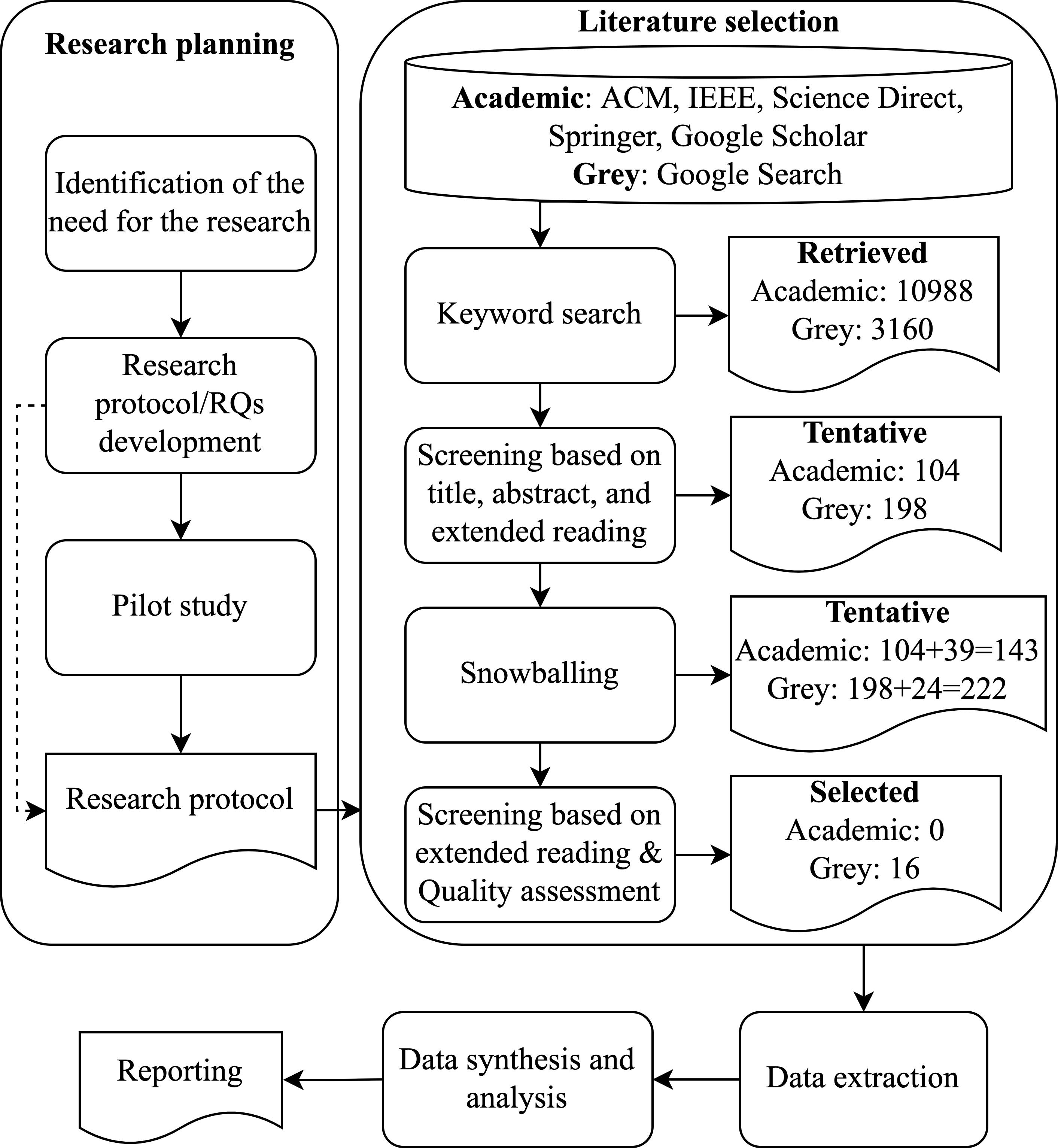}
\caption{Methodology overview.} \label{fig:methodology}
\end{figure}

\begin{itemize}
    \item RQ1: What are the characteristics of the existing AI risk assessment frameworks?
    \begin{itemize}
        \item RQ1.1 What are the demographics of the frameworks?
        \item RQ1.2 What RAI principles are addressed?
        \item RQ1.3 Who are the stakeholders?
        \begin{itemize}
            \item RQ1.3.1 Who conducts the assessment?
            \item RQ1.3.2 Whose activities are assessed?
        \end{itemize}
        \item RQ1.4 What is the scope of the frameworks?
        \begin{itemize}
            \item RQ1.4.1 Which development stages are covered by the frameworks?
            \item RQ1.4.2 Where can the frameworks be applied?
            \item RQ1.4.3 Which domains/sectors are the frameworks designed for?
        \end{itemize}
    \end{itemize}
    \item RQ2: How are the AI risks assessed?
    \begin{itemize}
        \item RQ2.1: What are the inputs?
        \item RQ2.2: What is the assessment process?
        \item RQ2.3: What are the outputs?
    \end{itemize}
\end{itemize}

n November 2022, we conducted a literature search using ACM, IEEE, Science Direct, Springer, Google scholar for academic papers and Google Search for industrial frameworks. The search terms used were \textit{(``artificial intelligence" OR ``machine learning" OR AI OR ML) AND (impact OR risk) AND (assess OR assessment OR assessing OR evaluate OR evaluation OR evaluating OR measure OR measurement OR measuring OR mitigate OR mitigation OR mitigating OR manage OR managing OR management)}. We only considered frameworks that have relatively concrete AI risk assessment solutions and exclude those higher-level ones.

We excluded academic papers from our study as we aimed to include frameworks currently being used in practice and academic papers either discuss identified industrial frameworks or lack details on AI risk assessment solutions. In total, we selected 16 industrial frameworks.
The complete research protocol is available \href{https://docs.google.com/document/d/1F_sAmRI7zvJBYyiF96cn5oNtG3O8psyB/edit?usp=sharing&ouid=111846093034327217492&rtpof=true&sd=true}{online}\footnote{\url{https://docs.google.com/document/d/1F_sAmRI7zvJBYyiF96cn5oNtG3O8psyB/edit?usp=sharing&ouid=111846093034327217492&rtpof=true&sd=true}}. 
We adopt \href{https://www.industry.gov.au/publications/australias-artificial-intelligence-ethics-framework/australias-ai-ethics-principles}{Australia's AI ethics principles} in this study: Human, societal and environmental (HSE) wellbeing, Human-centered values, Fairness, Privacy protection and security, Reliability and safety, Transparency and explainability, Contestability, and Accountability.


%% file: 3Results.tex
This section presents the results and findings of each RQ.

\subsection{RQ1: What are the characteristics of the existing AI risk assessment frameworks?}
This subsection discusses the characteristics of the collected frameworks based on the following aspects: demographics, RAI principles, stakeholders, software development lifecycle stages, geographical locations, and targeted sectors. We also present our findings and insights for some RQs.

To improve presentation, we first classify the frameworks based on whether they have clear specifications on different characteristics (e.g., whether RAI principles/stakeholders/stages are specified). Then, we further categorize them to see whether they have formulated the assessment and mitigation based on different sub-categories of those characteristics (e.g., different RAI principles).

\subsubsection{RQ1.1: What are the demographics of the frameworks?}

\begin{figure}
  \begin{subfigure}{0.5\columnwidth}
    \includegraphics[width=\linewidth]{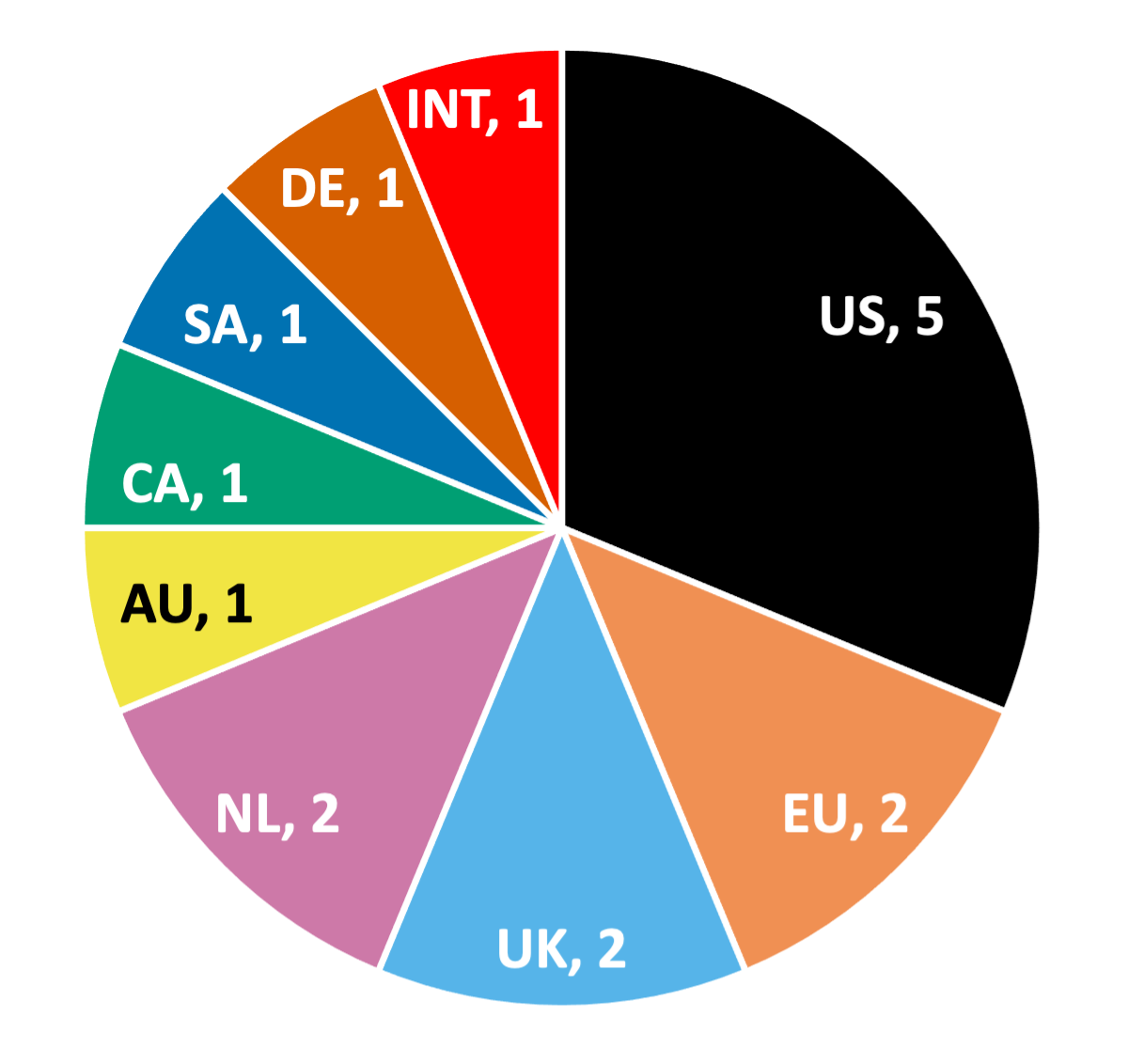}
    \caption{Number by region.} \label{fig:IndustryRegion}
  \end{subfigure}%
  \hspace*{\fill}   
  \begin{subfigure}{0.5\columnwidth}
    \includegraphics[width=\linewidth]{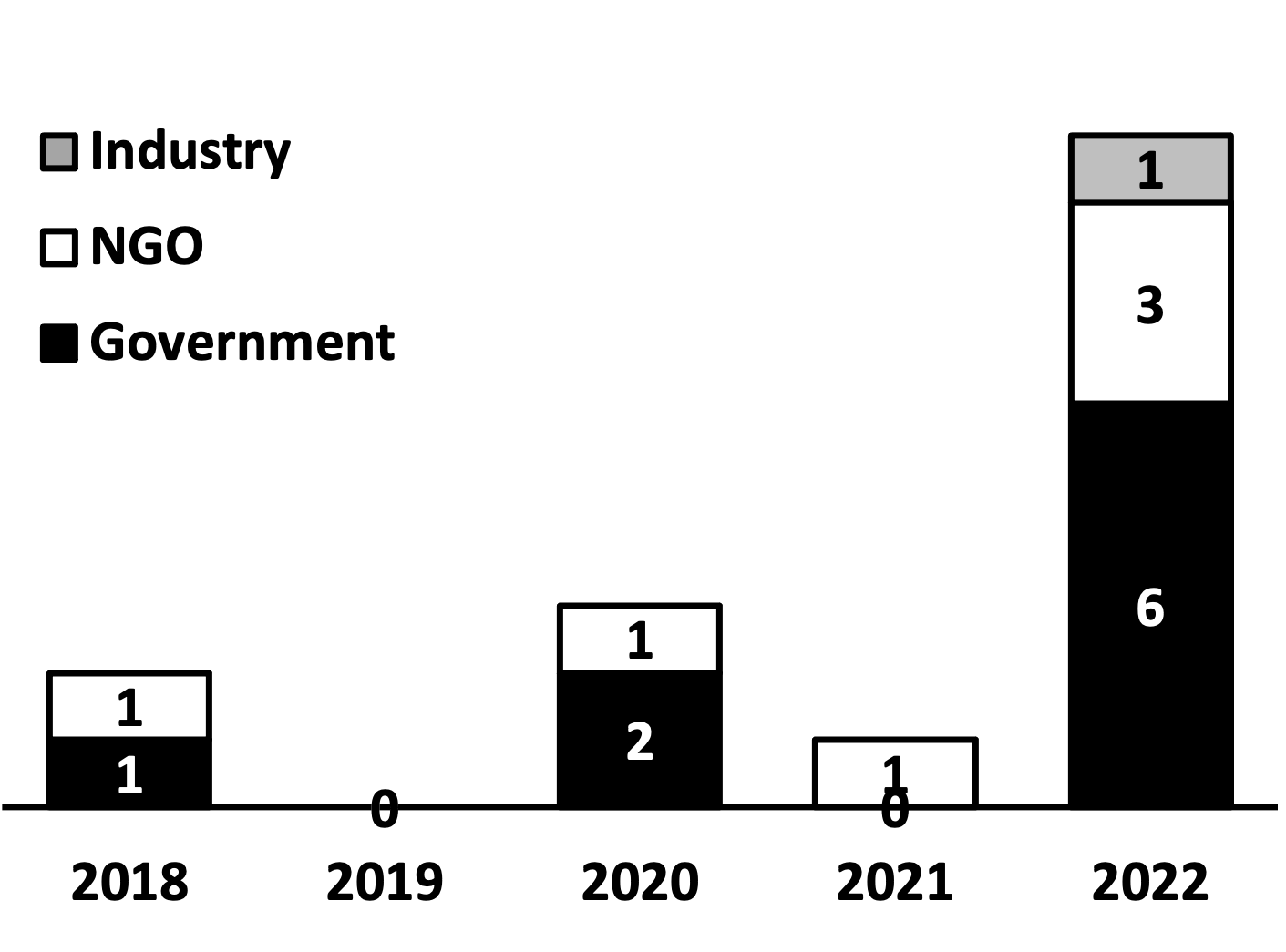}
    \caption{Number by year.} \label{fig:IndustryYearly}
  \end{subfigure}
    \begin{subfigure}{\columnwidth}
    \includegraphics[width=\linewidth]{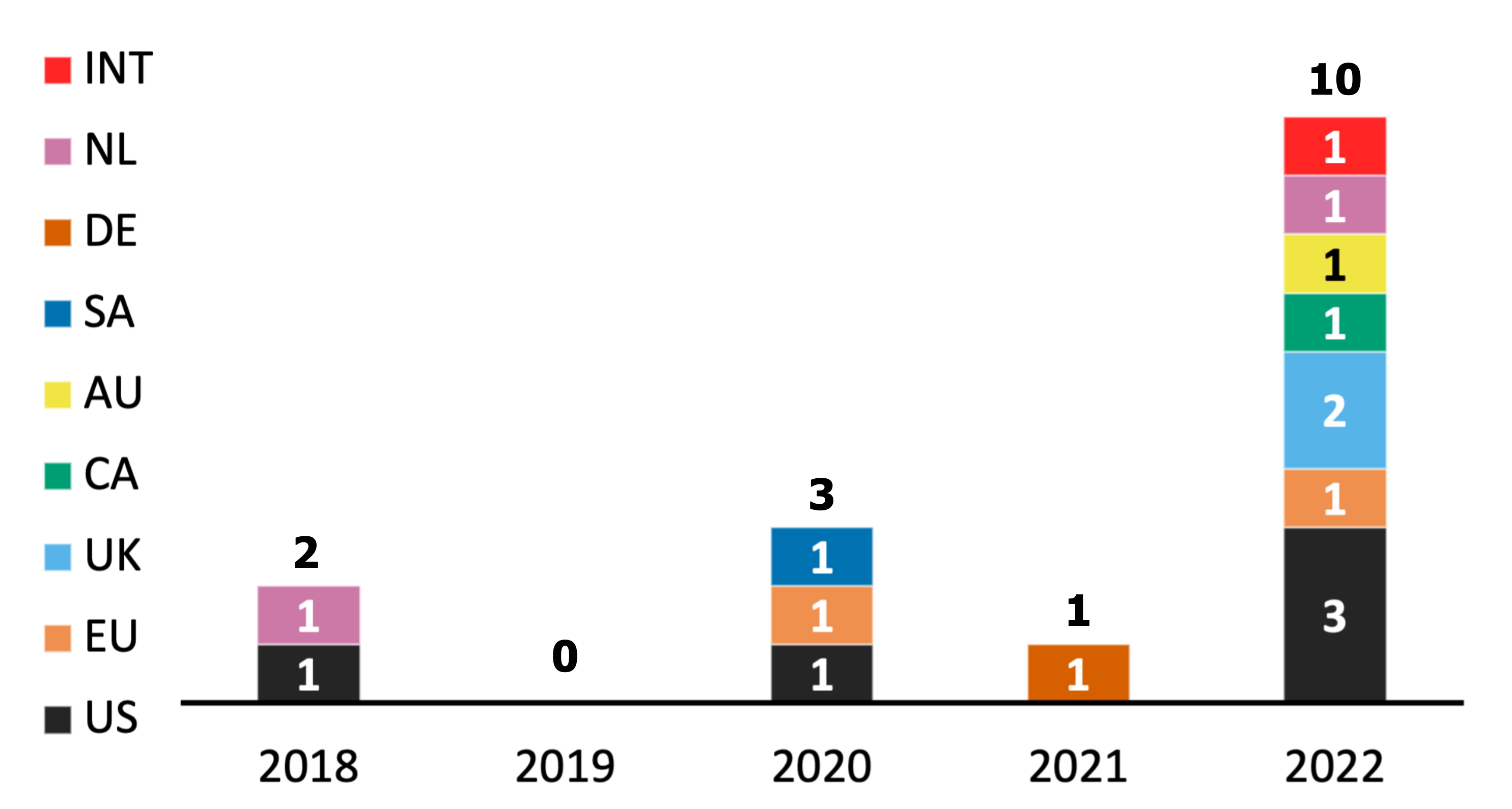}
    \caption{Number by year and region.} \label{fig:IndustryYearRegion}
  \end{subfigure}
\caption[test]{Demographics of collected frameworks.} \label{fig:IndustryDemo}
\end{figure}

As illustrated in Table \ref{tab:industrialFrameworks} and Fig. \ref{fig:IndustryDemo}, we identified 16 frameworks to be included in this study.
These frameworks have been published by organizations based in the United States (US), the United Kingdom (UK), the European Union (EU), Canada (CA), Australia (AU), Singapore (SA), the Netherlands (NL), and Germany (DE). Additionally, one framework has been released by the World Economic Forum, an international (INT) organization. 
Although we did not set a specific time limit for the literature search, as shown in Fig. \ref{fig:IndustryYearly}, the majority of the frameworks (10 out of 16, 62.5\%) were published or last updated (some frameworks tend to be updated over time) in 2022, followed by 1 framework (6.25\%) published in 2021, 3 frameworks (18.75\%) published in 2020, 0 framework published in 2019, and 2 (12.5\%) frameworks published in 2018.
Also, a significant proportion of the frameworks (9 out of 16, 56.25\%) were published by government agencies worldwide, with 6 of them last updated in 2022.
This suggests that the issue of AI risks has been gaining significant attention worldwide, particularly among government agencies.

\begin{sidewaystable*}[htpb]\fontsize{6.5pt}{0.55\baselineskip}\selectfont
\setlength{\tabcolsep}{2pt}
\caption{Industrial AI risk assessment frameworks (collected in November 2022).}
\label{tab:industrialFrameworks}
\begin{tabular}{clclccclccccccc}
\hline
\multicolumn{1}{l|}{} &
  \multicolumn{1}{c|}{} &
  \multicolumn{5}{c|}{\textbf{Demographics}} &
  \multicolumn{5}{c|}{\textbf{Characteristics}} &
  \multicolumn{3}{c}{\textbf{Processes}} \\ \cline{3-15} 
\multicolumn{1}{l|}{\multirow{-2}{*}{\textbf{No.}}} &
  \multicolumn{1}{c|}{\multirow{-2}{*}{\textbf{Frameworks}}} &
  \textbf{Region} &
  \multicolumn{1}{c}{\textbf{Affiliation}} &
  \textbf{Affiliation type} &
  \textbf{First release} &
  \multicolumn{1}{c|}{\textbf{Last update}} &
  \multicolumn{1}{c}{\textbf{RAI Princples}} &
  \textbf{Stakeholders} &
  \textbf{Stages} &
  \textbf{Region} &
  \multicolumn{1}{c|}{\textbf{Sector}} &
  \textbf{Type} &
  \textbf{Mitigation} &
  \textbf{*Risk factors} \\ \hline
\rowcolor[HTML]{EFEFEF} 
I1 &
  \begin{tabular}[c]{@{}l@{}}\href{https://www.nist.gov/system/files/documents/2022/08/18/AI_RMF_2nd_draft.pdf}{AI risk management framework}\\ (AI RMF)\cite{NIST_AIRMF}\end{tabular} &
  US &
  \begin{tabular}[c]{@{}l@{}}National Institute of Standards\\ and technology (NIST)\end{tabular} &
  Government &
  2022.05 &
  2022.08 &
  All principles &
  Specified &
  All stages &
  \begin{tabular}[c]{@{}c@{}}Region-\\ agnostic\end{tabular} &
  \begin{tabular}[c]{@{}c@{}}Sector-\\ agnostic\end{tabular} &
  Descriptive &
  *Yes &
  \begin{tabular}[c]{@{}c@{}}Hazard,\\ exposure,\\ vulnerability\end{tabular} \\
I2 &
  \begin{tabular}[c]{@{}l@{}}\href{https://altai.insight-centre.org/}{Assessment list for trustworthy AI}\\ (ALTAI)\cite{EUALTAI}\end{tabular} &
  EU &
  \begin{tabular}[c]{@{}l@{}}European Commission High-\\ Level Expert Group on AI\end{tabular} &
  Government &
  2019.06 &
  2020.07 &
  All principles &
  Specified &
  All stages &
  \begin{tabular}[c]{@{}c@{}}Region-\\ agnostic\end{tabular} &
  \begin{tabular}[c]{@{}c@{}}Sector-\\ agnostic\end{tabular} &
  Procedural &
  Yes &
  \begin{tabular}[c]{@{}c@{}}Hazard,\\ exposure,\\ vulnerability\end{tabular} \\
\rowcolor[HTML]{EFEFEF} 
I3 &
  \begin{tabular}[c]{@{}l@{}}\href{https://www.canada.ca/en/government/system/digital-government/digital-government-innovations/responsible-use-ai/algorithmic-impact-assessment.html}{Algorithm Impact Assessment}\\ tool (AIA)\cite{CAAIA}\end{tabular} &
  CA &
  Government of Canada &
  Government &
  2019 &
  2022.11 &
  Not specified &
  Not specified &
  \begin{tabular}[c]{@{}c@{}}Planning \&\\ requirements\\ analysis,\\ design, testing\end{tabular} &
  \begin{tabular}[c]{@{}c@{}}Region-\\ agnostic\end{tabular} &
  \begin{tabular}[c]{@{}c@{}}Sector-\\ agnostic\end{tabular} &
  Procedural &
  Yes &
  \begin{tabular}[c]{@{}c@{}}Hazard,\\ exposure,\\ vulnerability\end{tabular} \\
I4 &
  \begin{tabular}[c]{@{}l@{}}\href{https://www.government.nl/documents/reports/2021/07/31/impact-assessment-fundamental-rights-and-algorithms}{Fundamental rights and algorithm}\\ impact assessment (FRAIA)\cite{NLFRAIA}\end{tabular} &
  NL &
  \begin{tabular}[c]{@{}l@{}}Ministry of the Interior and\\ Kingdom Relations (BZK)\end{tabular} &
  Government &
  2022.03 &
  N/A &
  Not specified &
  Specified &
  All stages &
  \begin{tabular}[c]{@{}c@{}}Region-\\ agnostic\end{tabular} &
  \begin{tabular}[c]{@{}c@{}}Public\\ sectors\end{tabular} &
  Procedural &
  *Yes &
  \begin{tabular}[c]{@{}c@{}}Hazard,\\ exposure,\\ vulnerability\end{tabular} \\
\rowcolor[HTML]{EFEFEF} 
I5 &
  \begin{tabular}[c]{@{}l@{}}\href{https://ico.org.uk/for-organisations/guide-to-data-protection/key-dp-themes/guidance-on-ai-and-data-protection/ai-and-data-protection-risk-toolkit/}{AI and data protection risk toolkit}\\ \cite{UKICO}\end{tabular} &
  UK &
  \begin{tabular}[c]{@{}l@{}}Information Commissioner's\\ Office (ICO)\end{tabular} &
  Government &
  2021 &
  2022.05 &
  \begin{tabular}[c]{@{}l@{}}HSE wellbeing, human-\\ centered values, fairness,\\ privacy protection \&\\ security, reliability \&\\ safety, transparency \&\\ explainability,\\ accountability\end{tabular} &
  Specified &
  All stages &
  \begin{tabular}[c]{@{}c@{}}Reusable\\ anywhere\\ with\\ adjustments\end{tabular} &
  \begin{tabular}[c]{@{}c@{}}Sector-\\ agnostic\end{tabular} &
  Procedural &
  No &
  \begin{tabular}[c]{@{}c@{}}Hazard,\\ exposure,\\ vulnerability\end{tabular} \\
I6 &
  \begin{tabular}[c]{@{}l@{}}\href{https://www.pdpc.gov.sg/help-and-resources/2020/01/model-ai-governance-framework}{Model AI governance framework}\\ \cite{SAPDPC}\end{tabular} &
  SA &
  \begin{tabular}[c]{@{}l@{}}Personal Data Protection\\ Commission (PDPC)\end{tabular} &
  Government &
  2019.01 &
  2020.01 &
  \begin{tabular}[c]{@{}l@{}}HSE wellbeing, human-\\ centered values, fairness,\\ transparency \&\\ explainability, reliability\\ \& safety\end{tabular} &
  Not specified &
  Not specified &
  \begin{tabular}[c]{@{}c@{}}Region-\\ agnostic\end{tabular} &
  \begin{tabular}[c]{@{}c@{}}Sector-\\ agnostic\end{tabular} &
  Descriptive &
  Yes &
  \begin{tabular}[c]{@{}c@{}}Hazard,\\ exposure,\\ vulnerability\end{tabular} \\
\rowcolor[HTML]{EFEFEF} 
I7 &
  \begin{tabular}[c]{@{}l@{}}\href{https://www.digital.nsw.gov.au/policy/artificial-intelligence/nsw-artificial-intelligence-assurance-framework}{NSW artificial intelligence}\\ assurance framework\cite{NSW}\end{tabular} &
  AU &
  NSW Government &
  Government &
  2022.03 &
  N/A &
  \begin{tabular}[c]{@{}l@{}}HSE wellbeing, human-\\ centered values, fairness,\\ privacy protection \&\\ security, reliability \&\\ safety, transparency \&\\ explainability,\\ accountability\end{tabular} &
  Specified &
  All stages &
  \begin{tabular}[c]{@{}c@{}}Reusable\\ anywhere\\ with\\ adjustments\end{tabular} &
  \begin{tabular}[c]{@{}c@{}}Sector-\\ agnostic\end{tabular} &
  Procedural &
  *Yes &
  \begin{tabular}[c]{@{}c@{}}Hazard,\\ exposure,\\ vulnerability,\\ \textbf{mitigation risk}\end{tabular} \\
I8 &
  \href{http://ethicstoolkit.ai/}{Ethics \& algorithms toolkit}\cite{GovEX} &
  US &
  \begin{tabular}[c]{@{}l@{}}GovEX, the City and County\\ of San Francisco, Harvard\\ DataSmart, and Data\\ Community DC\end{tabular} &
  \begin{tabular}[c]{@{}c@{}}Government \\ involved\end{tabular} &
  2018 &
  N/A &
  \begin{tabular}[c]{@{}l@{}}HSE wellbeing, human-\\ centered values, fairness,\\ privacy protection\\ \& security, reliability \&\\ safety, transparency \&\\ explainability,\\ accountability\end{tabular} &
  Specified &
  Not specified &
  \begin{tabular}[c]{@{}c@{}}Region-\\ agnostic\end{tabular} &
  \begin{tabular}[c]{@{}c@{}}Sector-\\ agnostic\end{tabular} &
  Procedural &
  No &
  \begin{tabular}[c]{@{}c@{}}Hazard,\\ exposure,\\ vulnerability\end{tabular} \\
\rowcolor[HTML]{EFEFEF} 
I9 &
  \begin{tabular}[c]{@{}l@{}}\href{https://www.oa.pa.gov/Policies/Documents/rfd_bus012a.xlsx}{RFD-BUS012A} artificial\\ intelligence assessment tool\cite{RFDBUS012A}\end{tabular} &
  US &
  \begin{tabular}[c]{@{}l@{}}Pennsylvania Office of\\ Administration\end{tabular} &
  Government &
  2018.09 &
  2022.08 &
  Not specified &
  Not specified &
  \begin{tabular}[c]{@{}c@{}}Planning \&\\ requirements\\ analysis, design\end{tabular} &
  \begin{tabular}[c]{@{}c@{}}Region-\\ agnostic\end{tabular} &
  \begin{tabular}[c]{@{}c@{}}Sector-\\ agnostic\end{tabular} &
  Procedural &
  No &
  Vulnerability \\
I10 &
  \begin{tabular}[c]{@{}l@{}}\href{https://www.europeanlawinstitute.eu/fileadmin/user_upload/p_eli/Publications/ELI_Model_Rules_on_Impact_Assessment_of_ADMSs_Used_by_Public_Administration.pdf}{Model rules on impact assessment}\\ of algorithmic decision-making\\ systems used by public\\ administration\cite{EUELI}\end{tabular} &
  EU &
  \begin{tabular}[c]{@{}l@{}}European Law Institute\\ (ELI)\end{tabular} &
  NGO &
  2022.01 &
  N/A &
  Not specified &
  Not specified &
  Not specified &
  \begin{tabular}[c]{@{}c@{}}Region-\\ agnostic\end{tabular} &
  \begin{tabular}[c]{@{}c@{}}Public\\ sectors\end{tabular} &
  Procedural &
  Yes &
  \begin{tabular}[c]{@{}c@{}}Hazard,\\ exposure,\\ vulnerability\end{tabular} \\
\rowcolor[HTML]{EFEFEF} 
I11 &
  \begin{tabular}[c]{@{}l@{}}\href{https://www3.weforum.org/docs/WEF_Artificial_Intelligence_for_Children_2022.pdf}{Artificial intelligence for children}\\ toolkit\cite{WEF}\end{tabular} &
  INT &
  \begin{tabular}[c]{@{}l@{}}World Economic Forum\\ (WEF)\end{tabular} &
  NGO &
  2022.03 &
  N/A &
  \begin{tabular}[c]{@{}l@{}}HSE wellbeing, human-\\ centered values, fairness,\\ privacy protection \&\\ security, reliability \&\\ safety, transparency \&\\ explainability,\\ contestability\end{tabular} &
  Specified &
  Not specified &
  \begin{tabular}[c]{@{}c@{}}Region-\\ agnostic\end{tabular} &
  \begin{tabular}[c]{@{}c@{}}Children\\ \& youth\end{tabular} &
  Procedural &
  Yes &
  \begin{tabular}[c]{@{}c@{}}Hazard,\\ exposure,\\ vulnerability\end{tabular} \\
I12 &
  \begin{tabular}[c]{@{}l@{}}\href{https://ieeexplore.ieee.org/document/9084219}{Recommended practices} for\\ assessing the impact of\\ autonomous and intelligent\\ systems on human well-being\cite{IEEE_recommendations}\end{tabular} &
  US &
  IEEE &
  NGO &
  2020.05 &
  N/A &
  \begin{tabular}[c]{@{}l@{}}HSE wellbeing, human-\\ centered values\end{tabular} &
  Specified &
  All stages &
  \begin{tabular}[c]{@{}c@{}}Region-\\ agnostic\end{tabular} &
  \begin{tabular}[c]{@{}c@{}}Sector-\\ agnostic\end{tabular} &
  Descriptive &
  *Yes &
  \begin{tabular}[c]{@{}c@{}}Hazard,\\ exposure,\\ vulnerability\end{tabular} \\
\rowcolor[HTML]{EFEFEF} 
I13 &
  \begin{tabular}[c]{@{}l@{}}\href{https://blogs.microsoft.com/wp-content/uploads/prod/sites/5/2022/06/Microsoft-RAI-Impact-Assessment-Template.pdf}{Responsible AI impact assessment}\\ template\cite{MS_RAItemplate}\end{tabular} &
  US &
  Microsoft &
  Industry &
  2022.06 &
  N/A &
  All principles &
  Not specified &
  Not specified &
  \begin{tabular}[c]{@{}c@{}}Region-\\ agnostic\end{tabular} &
  \begin{tabular}[c]{@{}c@{}}Sector-\\ agnostic\end{tabular} &
  Procedural &
  No &
  \begin{tabular}[c]{@{}c@{}}Hazard,\\ exposure,\\ vulnerability\end{tabular} \\
I14 &
  \begin{tabular}[c]{@{}l@{}}\href{https://www.adalovelaceinstitute.org/project/algorithmic-impact-assessment-healthcare/}{Algorithmic impact assessments}\\ (AIAs) in healthcare\cite{AdaLovelace}\end{tabular} &
  UK &
  Ada Lovelace Institute &
  NGO &
  2022.01 &
  N/A &
  \begin{tabular}[c]{@{}l@{}}HSE wellbeing, human\\ centered values, fairness,\\ privacy protection \&\\ security,  reliability \&\\ safety, transparency \&\\ explainability\end{tabular} &
  Specified &
  \begin{tabular}[c]{@{}c@{}}Planning \&\\ requirements\\ analysis,\\ design, tetsing,\\ deployment,\\ monitoring\end{tabular} &
  UK only &
  \begin{tabular}[c]{@{}c@{}}UK\\ healthcare\end{tabular} &
  Procedural &
  *Yes &
  \begin{tabular}[c]{@{}c@{}}Hazard,\\ exposure,\\ vulnerability\end{tabular} \\
\rowcolor[HTML]{EFEFEF} 
I15 &
  \begin{tabular}[c]{@{}l@{}}\href{https://ecp.nl/publicatie/artificial-intelligence-impact-assessment-english-version/}{Artificial intelligence impact}\\ assessment\cite{ECP}\end{tabular} &
  NL &
  \begin{tabular}[c]{@{}l@{}}ECP, Platform for the\\ Information Society\end{tabular} &
  NGO &
  2018 &
  N/A &
  Not specified &
  Specified &
  Not specified &
  \begin{tabular}[c]{@{}c@{}}Region-\\ agnostic\end{tabular} &
  \begin{tabular}[c]{@{}c@{}}Sector-\\ agnostic\end{tabular} &
  Procedural &
  *Yes &
  \begin{tabular}[c]{@{}c@{}}Hazard,\\ exposure,\\ vulnerability,\\ \textbf{mitigation risk}\end{tabular} \\
I16 &
  \begin{tabular}[c]{@{}l@{}}\href{https://uploads-ssl.webflow.com/5f57d40eb1c2ef22d8a8ca7e/61a8efeb70b677a50f9c24cb_2021_AW_Decision_Public_Sector_EN_v5.pdf}{Automated decision-making}\\ systems in the public sector: an\\ impact assessment\\ tool for public authorities\cite{AlgoWatch}\end{tabular} &
  DE &
  Algorithm Watch &
  NGO &
  2021.06 &
  N/A &
  All principles &
  Not specified &
  Not specified &
  \begin{tabular}[c]{@{}c@{}}Region-\\ agnostic\end{tabular} &
  \begin{tabular}[c]{@{}c@{}}Public\\ sectors\end{tabular} &
  Procedural &
  *Yes &
  \begin{tabular}[c]{@{}c@{}}Hazard,\\ exposure,\\ vulnerability\end{tabular} \\ \hline
\end{tabular}
\end{sidewaystable*}

In terms of the number of published frameworks (Fig. \ref{fig:IndustryRegion} and Fig. \ref{fig:IndustryYearRegion}), the US leads in research and development on AI risk assessment and published 6 related frameworks, including 3 frameworks from US government agencies (i.e., I1 by National Institute of Standards and Technology (NIST), and I8 and I9 by two state/city government agencies) and 2 from US-based organizations (i.e., I12 by IEEE and I13 by Microsoft).
The UK, EU, and Netherlands rank second with 2 frameworks developed.
In the UK, one framework was developed by a government agency (Information Commissioner's Office, ICO) and Ada Lovelace Institute published the other specifically for the proposed National Medical Imaging Platform of the National Health Service (NHS).
The EU published 2 frameworks on AI risk assessment in 2020 (last updated) and 2022, respectively. Notably, the EU has drafted its \href{https://artificialintelligenceact.eu/}{AI Act}, which marks a significant step towards operationalizing RAI by legislation.
The Netherlands published 2 frameworks, one in 2018 by a non-government organization (ECP) and the other in 2022 by its Ministry of the Interior and Kingdom Relations (BZK).
Australia's New South Wales government published the nation's first AI Assurance Framework in 2022.
Singapore had its AI governance framework published in 2019, while it launched AI Verify in May 2022 to objectively assess AI systems in a verifiable way. The Canadian government released its Algorithm impact assessment tool in 2021.
The World Economic Forum (WEF) published a toolkit for managing AI risks to children.
Lastly, one framework published by a German-based organization, Algorithm Watch, is identified in this study.

\begin{center}
\begin{tcolorbox}[breakable, colback=gray!10,
colframe=black, 
width=\columnwidth,
arc = 1mm,
boxrule=0.5 pt,
]
\textbf{Finding to RQ1.1}: 
The growing number of AI risk assessment frameworks worldwide indicates increasing global concern about the risks associated with the development and use of AI systems and a growing recognition of RAI approaches to assess and mitigate AI risks.


\end{tcolorbox}
\end{center}

\subsubsection{RQ1.2 What RAI principles are addressed?}
This sub-RQ aims to investigate the RAI principles (i.e., the corresponding risk category) addressed by the identified frameworks. We have mapped the various principles from different frameworks to \href{https://www.industry.gov.au/publications/australias-artificial-intelligence-ethics-framework/australias-ai-ethics-principles}{Australia's AI ethics principles} (see Table \ref{tab:industrialFrameworks}).

\begin{figure}
\centering
  \begin{subfigure}{0.85\columnwidth}
    \includegraphics[width=\linewidth]{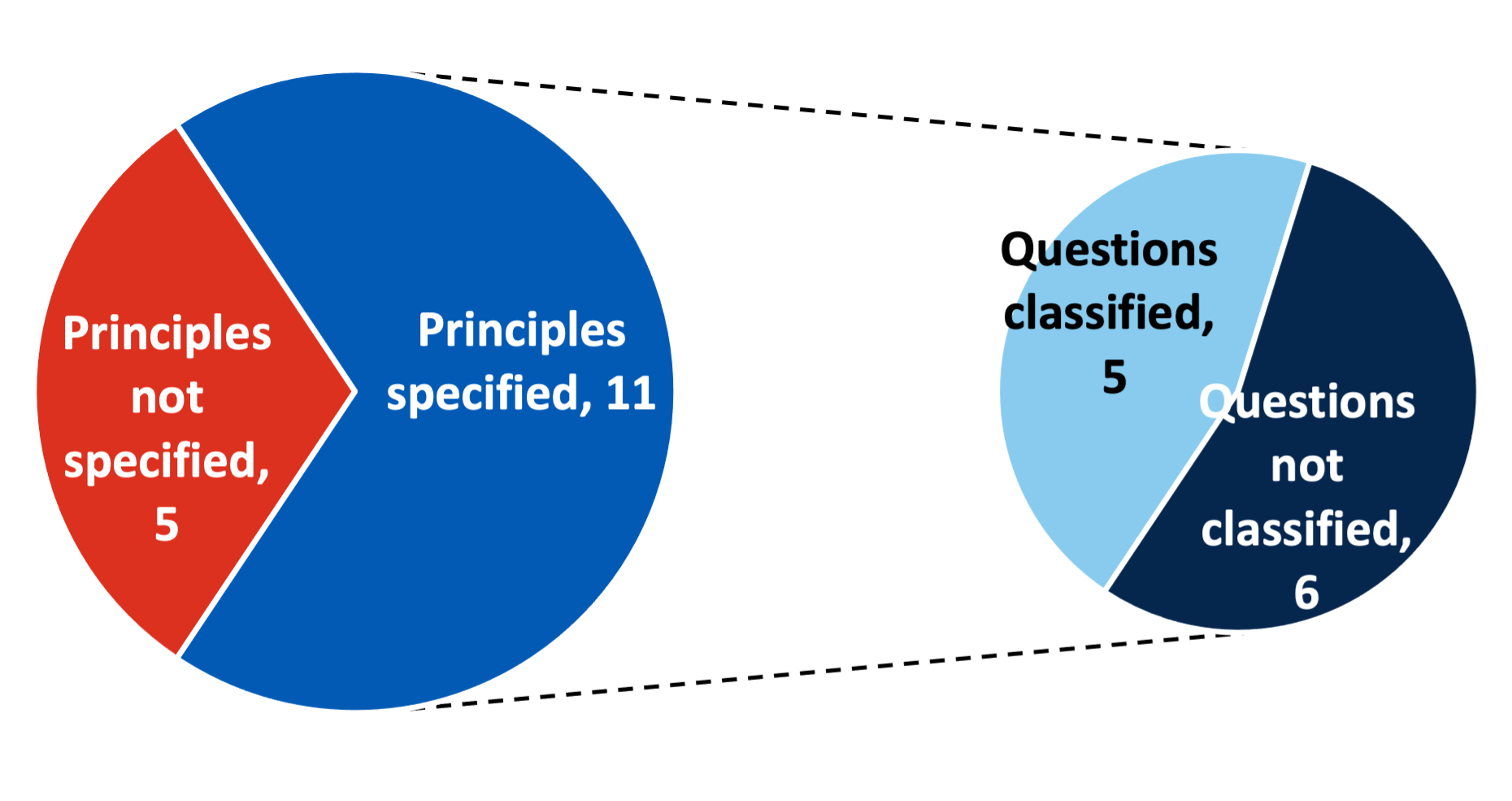}
    \caption{Principle overview.} \label{fig:principleOverview}
  \end{subfigure}%
  
    \begin{subfigure}{\columnwidth}
    \includegraphics[width=\linewidth]{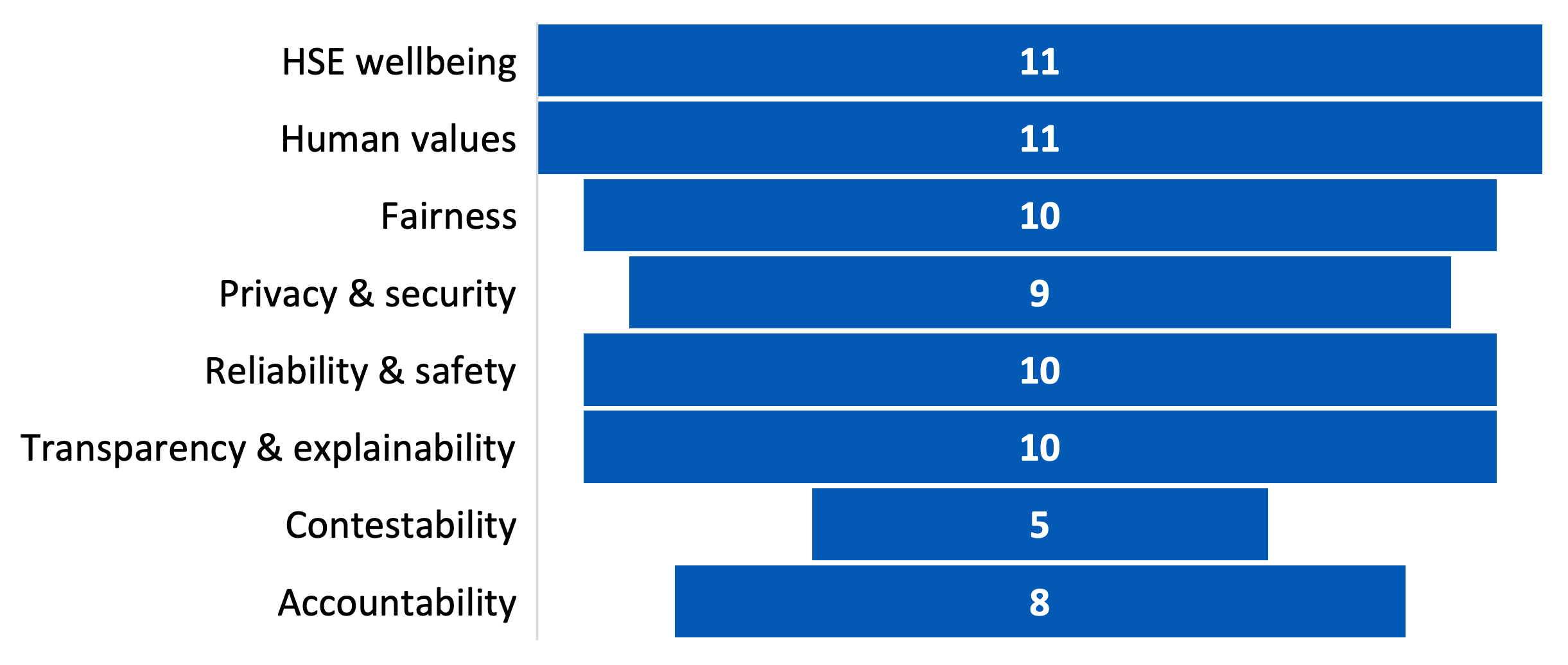}
    \caption{Principle coverage.} \label{fig:principlesStatis}
  \end{subfigure}
\caption[test]{RAI principles covered by the identified frameworks.} \label{fig:principles}
\end{figure}

As illustrated in Fig. \ref{fig:principleOverview}, among the 16 identified frameworks, 11 frameworks (I1, I2, I5-I8, I11-I14, I16) have specified their guiding principles.
5 frameworks (I3, I4, I9, I10, I15) do not explicitly state their corresponding principles, although they may implicitly encompass these principles through their framework description and introduction (e.g., I3) or references to other existing frameworks, standards, and guidelines (e.g., I4).
Among the 11 frameworks with specified guiding principles, only 5 frameworks (i.e., I1, I2, I11, I13, I16) organize their sets of AI risk assessment questions or checklists based on different RAI principles.

All the 11 frameworks that explicitly specify the guiding principles or targeted risks consider HSE wellbeing and human-centred values. Out of these 11 frameworks, 10 frameworks cover fairness, reliability \& safety, transparency \& explainability. The only exception is framework I12, which focuses mainly on HSE wellbeing.
Privacy protection \& security is covered by 9 (I1, I2, I5, I7, I8, I11, I13, I14, I16) and accountability is covered by 8 frameworks (I1, I2, I5, I7, I8, I11, I13, I16). Only 5 frameworks (I1, I2, I11, I13, I16) include contestability (Fig. \ref{fig:principlesStatis}).

\subsubsection{RQ1.2: Who are the stakeholders?}
\begin{figure}
\centering
\includegraphics[width=\columnwidth]{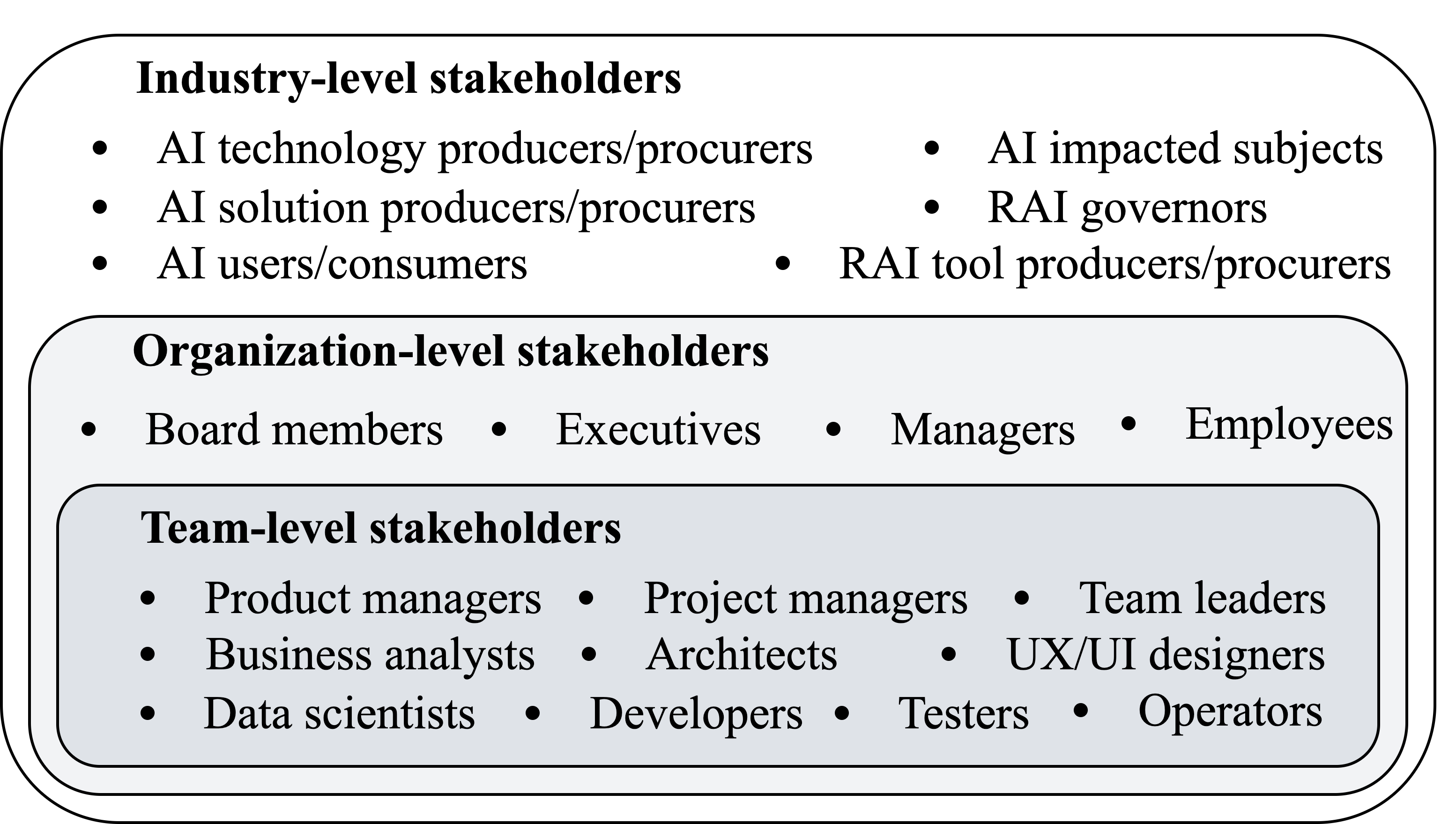}
\caption{Stakeholders classification\cite{lu2022responsible}.} \label{fig:stakeholderType}
\end{figure}

This subsection examines the stakeholders involved in the frameworks from two perspectives: the framework user(s) who are responsible for conducting the risk assessment (i.e., \textbf{assessor}), and those whose activities are being assessed (i.e., \textbf{assessee}).
The stakeholders classification is based on our previous study \cite{lu2022responsible}, where the stakeholders are categorized into three levels: industry-level, organization-level, and team-level (see Fig. \ref{fig:stakeholderType}).

\begin{figure}
\centering
\includegraphics[width=0.85\columnwidth]{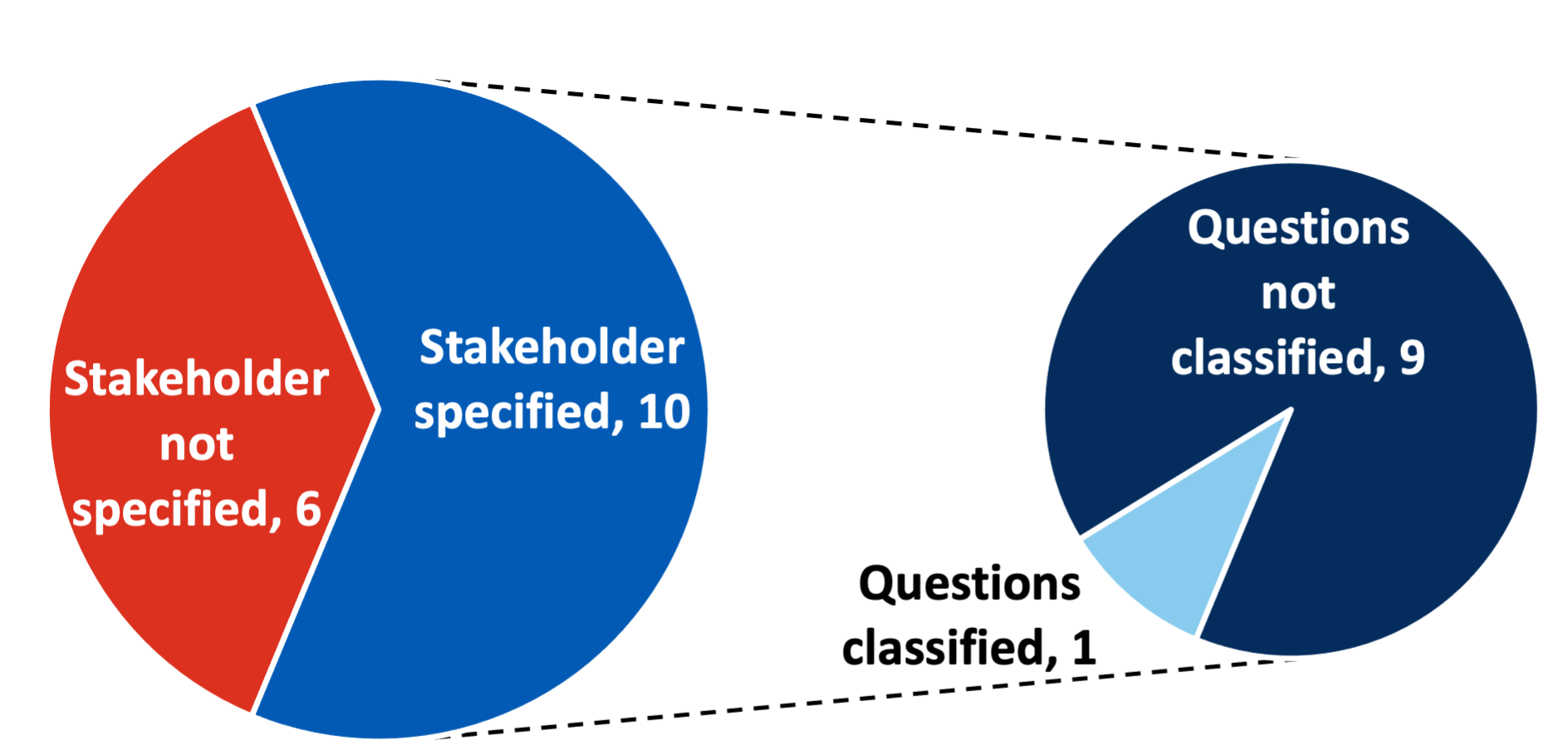}
\caption{Stakeholders of collected frameworks.} \label{fig:stakeholder}
\end{figure}

Fig. \ref{fig:stakeholder} shows that 10 of the collected frameworks (I1, I2, I4, I5, I7, I8, I11, I12, I14, I15) have mentioned their targeted stakeholders.
For example, NIST's AI RMF (I1) specifies the framework is intended for ``AI actors" defined by the Organisation for Economic Co-operation and Development (OECD), while EU's ALTAI (I2) has listed the example stakeholders in its guide on ``\href{https://altai.insight-centre.org/Home/HowToComplete}{How to complete ALTAI}\footnote{\url{https://altai.insight-centre.org/Home/HowToComplete}}".
However, only the Netherlands BZK's FRAIA (I4) has clearly specified the different stakeholders associated with different assessment stages to answer stage-specific questions.

The data synthesis results show that
all 16 frameworks involve the participation of RAI governors as the assessors and development teams (e.g., data scientists, system developers) as the assessee.
RAI governors are those who set and enforce RAI policies within an organization or community, and they can be internal or external.

One issue identified through the data synthesis process is the lack of consideration of more diverse and inclusive (i.e., comprehensive) roles of stakeholders from different levels.
For example, industry-level procurers are largely neglected, with only I1, I2 and I7 considering this aspect.
Team-level speaking, all 10 frameworks with identified stakeholders require input from AI system development teams (i.e., assessees) on information such as intended use, data source, data privacy, and algorithm transparency.
The assessees typically include product managers, project managers, team leaders, data scientists, and system developers. 
However, 9 out of 10 frameworks fail to consider more diverse roles of assessees (e.g., architects, UI/UX designers\cite{NIST_AIRMF, lu2022responsible}).
The US NIST's AI RMF (I1) is distinguished by its inclusion of a broader range of stakeholders involved in various stages of AI system development and post-development, such as procurement, deployment, and operations. However, I1 does not  explicitly present categorized assessments and mitigations based on different stakeholders.

\begin{center}
\begin{tcolorbox}[breakable, colback=gray!10,
colframe=black, 
width=\columnwidth,
arc = 1mm,
boxrule=0.5 pt,
]
\textbf{Finding to RQ1.2 \& RQ1.3}: 
Many of the collected AI risk assessment frameworks have a limited scope and fail to consider important factors such as more comprehensive stakeholders and RAI principles. This can result in certain risks being overlooked and left unaddressed. 
\end{tcolorbox}
\end{center}

\subsubsection{RQ1.4 What is the scope of the frameworks?}
With the RQ, we aim to explore the scope of the existing AI risk assessment frameworks. 

a) RQ1.4.1: Which development stages are covered by the frameworks?

\begin{figure*}[]
\centering
\includegraphics[width=\textwidth]{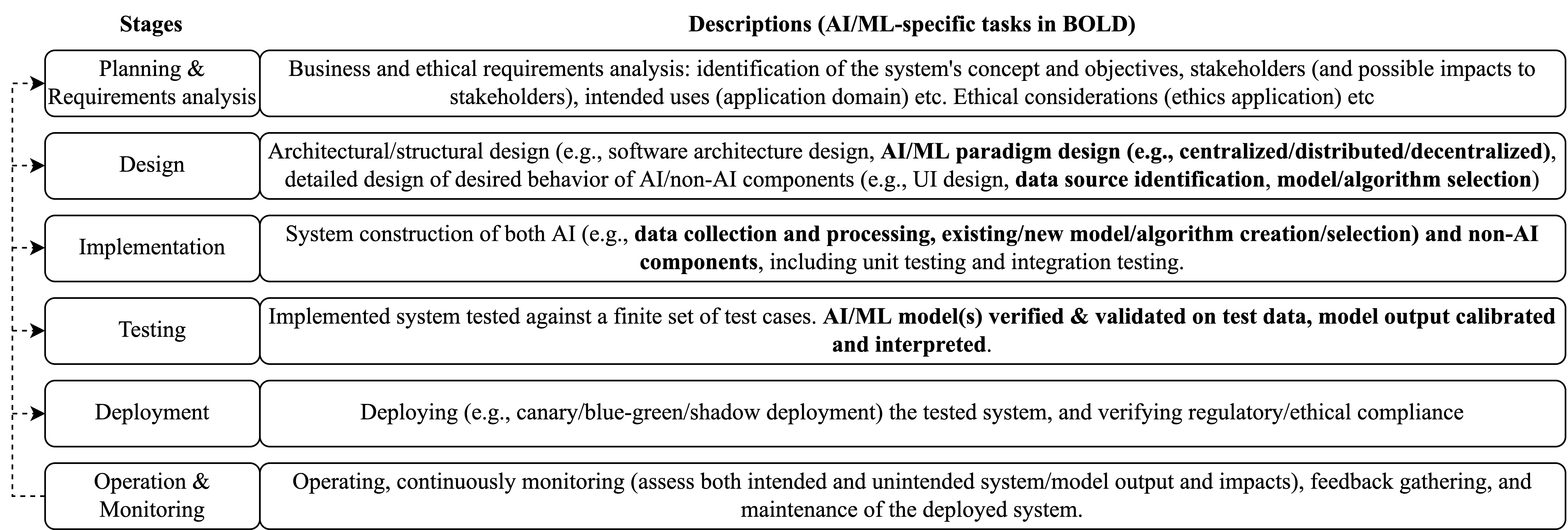}
\caption{AI system lifecycle (adapted from\cite{NIST_AIRMF, amershi2019software, 10.1145/3522664.3528607, UKICO}).} \label{fig:sdlc}
\end{figure*}

This RQ aims to investigate the stages covered by the collected frameworks in the AI system development lifecycle (AI-SDLC).
By referencing several existing sources with AI-SDLC \cite{NIST_AIRMF, amershi2019software, 10.1145/3522664.3528607, UKICO}, we first summarized the typical stages included in (AI) SDLC (i.e., planning \& requirement analysis, design, implementation, testing, deployment, operation \& monitoring). Then, we adapted the tasks in each stage with additional AI-specific context and derived an AI-SDLC (Fig. \ref{fig:sdlc}).
The detailed results of the AI-SDLC stages covered by the collected frameworks are presented in Table \ref{tab:industrialFrameworks}.

\begin{figure}
\centering
\includegraphics[width=0.85\columnwidth]{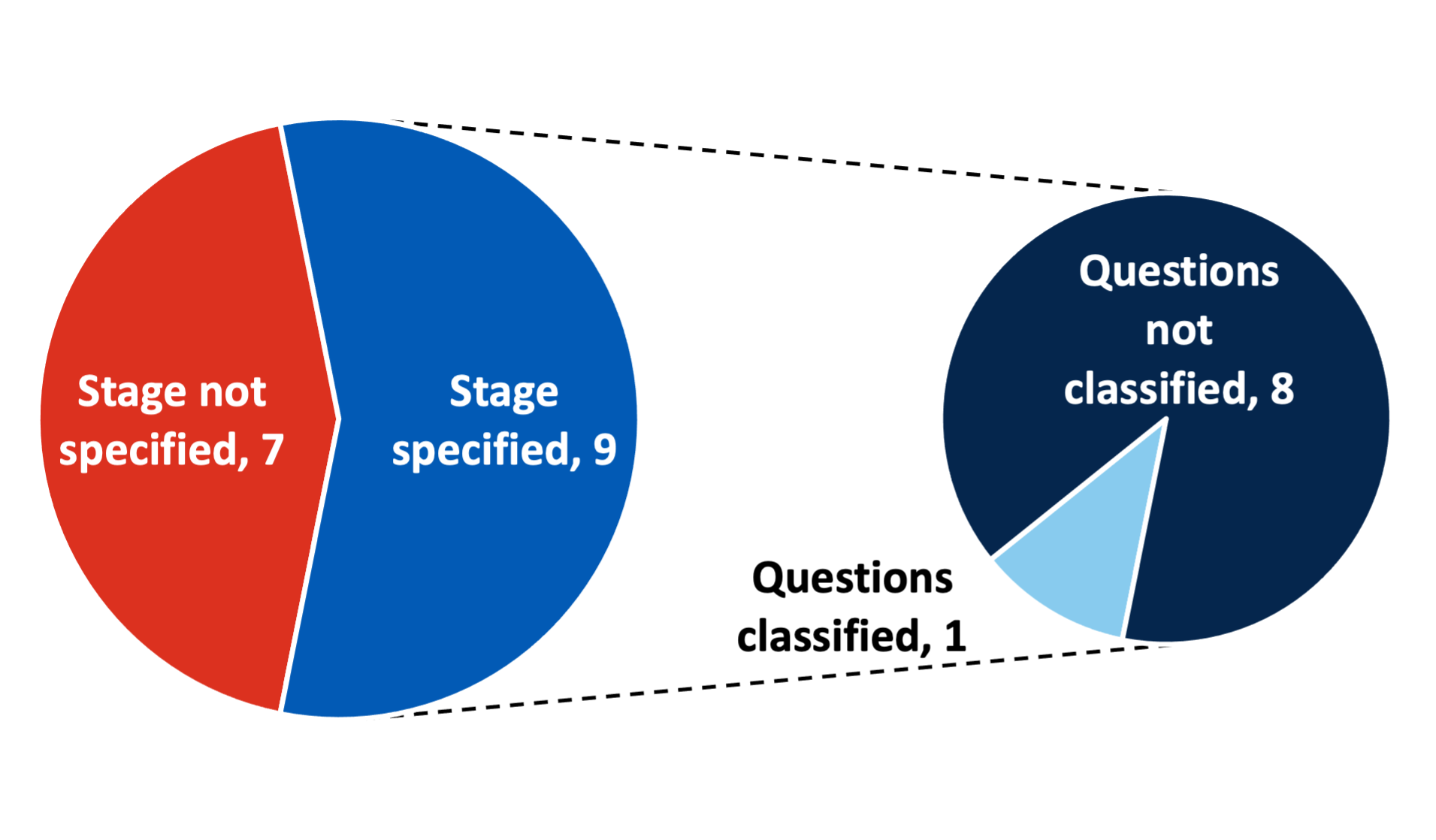}
\caption{Stages covered by collected frameworks.} \label{fig:stageOverview}
\end{figure}

Fig. \ref{fig:stageOverview} shows that 7 of the collected frameworks do not specify AI system lifecycle stages.
Although the other 9 frameworks (I1-I5, I7, I9, I12, I14) have clarified when they can be applied during the AI system lifecycle, The UK ICO's AI and Data Protection Risk toolkit (I5) is the only one that has categorized AI risk assessment and evaluation processes based on different stages of the AI system lifecycle.
Netherlands BZK's FRAIA (I4) is similarly structured in a more coarse-grained way in that the assessment is conducted based on three stages: input, throughput, and output. 

6 (I1, I2, I4, I5, I7, I12) out of 9 frameworks with specified AI system lifecycle stages can be used to evaluate potential risks throughout the entire AI system lifecycle.
The other 3 frameworks (I3, I9, I14) focus on the initial stages of ideation (i.e., planning \& requirements analysis, design).
In addition, I3 covers the testing stage, while I14 covers the testing, deployment and follow-up monitoring stages.

b) RQ2.3.2: Where can the frameworks be applied?

With the RQ, we aim to explore whether there are geographical constraints to applying the existing AI risk assessment frameworks.

The government-developed frameworks (i.e., I1-I9) can be applied anywhere, although some of them may require adjustments considering region-specific elements in the frameworks.
For example, The UK ICO's AI and Data Protection Risk toolkit (I5) is aligned with the UK's General Data Protection Regulation (GDPR), and the AU NSW's AI Assurance Framework (I7) references relevant policies in the Australian state of New South Wales.
6 out of 7 frameworks developed by NGOs and industrial companies (I10-I13, I15-I16) are region-agnostic. At the same time, I14 is specially designed for UK NHS's planned National Medical Imaging platform.

c) RQ1.4.2 Which domains/sectors are the frameworks designed for?

This RQ intends to investigate the domains and sectors where the frameworks can be applied.

Most of the collected frameworks are generally designed across various domains. However, 5 frameworks have been designed for specific purposes. FRAIA (I4), Model Rules (I10) and Impact Assessment Tool for Public Authorities (I16) are intended for evaluating the development and deployment of AI systems in the public sector. The AI for Children toolkit (I11) is specifically designed for AI systems that may impact children and youth as potential users. AIAs in Healthcare (I15) is intended to assess risks associated with designing and developing AI systems that require access to the UK National Medical Imaging Platform.

\begin{center}
\begin{tcolorbox}[breakable, colback=gray!10,
colframe=black, 
width=\columnwidth,
arc = 1mm,
boxrule=0.5 pt,
]
\textbf{Finding to RQ1.4}:
While the collected AI risk assessment frameworks are generally domain-agnostic and holistically consider the entire lifecycle of AI systems, they lack clear guidance on how to adapt them to diverse contexts. This limitation hinders the effectiveness of these frameworks, as risks and mitigation measures may vary depending on the specific context in which AI systems are used, such as different organizations, sectors or regions.
\end{tcolorbox}
\end{center}

\subsection{RQ2: How are AI risks assessed?} 
\label{Sec:RQ2}
This section presents the assessment processes of the collection frameworks.

The frameworks are categorized into two types: \textbf{procedural} and \textbf{descriptive}.
The descriptive frameworks are less concrete by providing general non-prescriptive assessment and mitigation and not referring to more specific and concrete solutions.
In contrast, procedural frameworks are more structured and include more detailed steps (e.g., inputs, processes, outputs) for conducting AI risk assessments. The procedural frameworks can also contain suggested mitigation solutions, assessment templates, or checklists.

The collected frameworks examine underlying risks and/or corresponding mitigation plans. 
To better present the results, we summarize the different types of risks (i.e., \textbf{risk factors}) the frameworks take into account. We adapted the risk categorization from a traditional risk management framework \cite{VIC_RMP} and added AI-specific context. The adapted risk factors are categorized as follows:
\begin{itemize}
    \item \textbf{Hazard}: A hazard refers to any dangerous situation or condition arising from AI systems or related activities/artifacts that can result in harm to HSE wellbeing. Hazards are sources of harm or exploit external to AI systems.
    \item \textbf{Exposure}: Exposure refers to individuals, property, systems, or other elements located within zones affected by AI-related hazards that are therefore at risk of potential losses.
    \item \textbf{Vulnerability}: Vulnerability pertains to the characteristics and circumstances of an AI system or related artifacts that make it susceptible to the detrimental effects of a hazard. Compared to hazards, vulnerabilities are internal weaknesses/issues of AI systems.
    \item \textbf{Risks by/after mitigation (Mitigation risk)}: Mitigation risks refer to the potential newly introduced risks brought about by the implementation of specific mitigation, resilience, or control measures, or residual risks that persist even after the implementation of mitigation measures.
\end{itemize}

For each of the collected frameworks, we summarized their types (i.e., descriptive or procedural) and examined mitigation measures and risk factors in Table \ref{tab:industrialFrameworks}.

We only articulate the answers to RQ3.1 (framework inputs) and RQ3.3 (framework outputs) for the procedural framework, as the descriptive frameworks do not have direct inputs or outputs. RQ3.2 (assessment processes) fits all frameworks, and the answer is thus presented based on all frameworks collected.

\subsubsection{RQ2.1: What are the inputs?}
This RQ investigate the inputs and the forms of inputs of the procedural frameworks.

The procedural frameworks are all based on certain forms of questionnaires (e.g., self-assessment template, checklist etc.). The inputs to these frameworks are answers to predefined questions provided by relevant stakeholders (e.g., development teams including system developers, data scientists etc.).

2 frameworks (EU ALTAI, I2 and CA AIA, I3) are designed as interactive online tools. Users can input the required information about their AI systems and get instant feedback based on their inputs.
Similarly, I5 and I9 are based on excel sheets where users can fill in system details or check if the recommended practices for minimizing potentials are met.
I13 and I14 provide self-assessment templates where predefined questions regarding the AI system (e.g., intended use, stakeholders, benefits/harms) need to be answered.
The other seven procedural frameworks (I4, I7, I8, I10, I11, I15, I16) are available as published reports, where more detailed descriptions of the contexts are given. In these reports, AI risk/impact assessment questionnaires/checklists are given.
It is important to note that in Q\&A-style assessments, both the quality of the answers and the underpinning methodology used to generate them are crucial factors, rather than relying solely on subjective inputs from the assessors.
\begin{center}
\begin{tcolorbox}[breakable, colback=gray!10,
colframe=black, 
width=\columnwidth,
arc = 1mm,
boxrule=0.5 pt,
]
\textbf{Finding to RQ2.1}: The current AI risk assessment frameworks primarily rely on subjective evaluation from assessors via a series of questions or checklists. This can lead to potentially biased results, as subjective evaluations can be influenced by the assessors' individual biases, beliefs, and experiences. To improve the accuracy and reliability of AI risk assessments, objective tools and techniques should be developed and incorporated into the risk assessment process.
\end{tcolorbox}
\end{center}

\subsubsection{RQ2.2: What are the processes?}
This section discusses how risk assessments are conducted in both descriptive and procedural frameworks.

The descriptive industrial frameworks include AI RMF (I1) by US NIST, Model AI governance framework (I6) by Singapore, and recommended practices for assessing the impact of autonomous and intelligent systems on human well-being (I12) by IEEE.

AI RMF (I1) is a framework with four components (map, measure, manage, and govern) that gives organizations recommendations to adopt and adapt to their specific needs.
AI RMF (I1) is a non-prescriptive framework that aims to identify, assess, and manage context-related risks by presenting desired outcomes and general approaches for risk management. It promotes the development of a culture of active risk management through its recommendations and non-exhaustive solutions presented in its \href{https://pages.nist.gov/AIRMF/}{companion playbook}\footnote{\url{https://pages.nist.gov/AIRMF/}}. AI RMF (I1) is a non-prescriptive framework that aims to identify, assess, and manage context-related risks by presenting desired outcomes and general approaches for risk management. It promotes the development of a culture of active risk management through its recommendations and non-exhaustive solutions presented in its companion playbook.
Similarly, Singapore's Model Framework (I6) and IEEE's standard on AI impact assessment (I12) are designed to be flexible by providing higher-level guidance on the assessment processes.

As for the procedural frameworks, the assessment processes are based on the input answers, where potential risks are identified through the Q\&A processes.
The assessment and evaluation processes of various procedural frameworks can be grouped into four categories: risk/principle-based (I2, I5, I7, I8, I11, I16), system development process-based (I4, I5), essential system component-based (I3, I9), and system description- and requirements-based (I10, I13, I14, I15).
The risk/principle-based assessments include questions designed for each of the different risks/principles.
The process-based assessments include questions throughout different AI-SDLC stages, from planning to monitoring \& operations.
The component-based assessments are formulated based on essential components (e.g., algorithms, data).
The system description- and requirements-based solutions offer mechanisms for the assessee to provide information about their AI systems and reflect on compliance with specific requirements.

For the more developed tools and frameworks, such as I2 and I3, the risk scores and potential risks are calculated automatically based on the selections/inputs.
As for other procedural frameworks, such as report- or excel-based ones, they identify and assess risks by the assessment conductors through a more manual process. The assessors evaluate the system's details, such as intended and unintended uses, stakeholders, data integrity, algorithmic explainability, and consult with external or internal stakeholders if necessary. This process enables a seemingly systematic analysis of an AI system to evaluate its impact and risks.

A valuable part of the AI risk assessment is the mitigation plans suggested by some frameworks. In Table \ref{tab:industrialFrameworks}, we summarize whether clear mitigation considerations are included in the frameworks by examining the questions/recommendations included in each of the 16 frameworks (\textit{Yes: Mitigation specified. *Yes: Mitigation included but not specified. No: Mitigation not included}).
Only 5 (I2, I3, I6, I10, I11) out of 16 frameworks have specified mitigation-related aspects.
7 frameworks (I1, I4, I7, I12, I14-I16) have more or less included risk mitigation measures without clearly specifying them.
4 frameworks (I5, I8, I9, I13) do not cover mitigation.

\begin{figure}
\centering
\includegraphics[width=0.85\columnwidth]{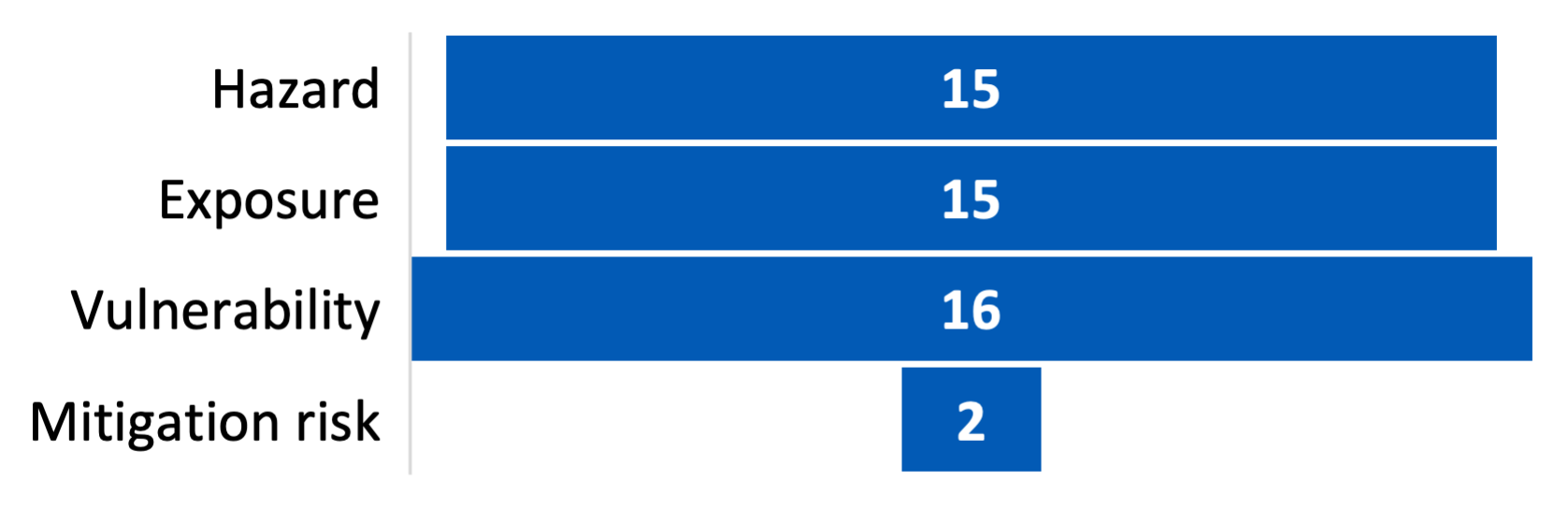}
\caption{Risk factors considered by collected frameworks.} \label{fig:Riskfactors}
\end{figure}

As for the risk factors, none of the frameworks specified the different risk factors they considered. However, given the potential value of such categorization in helping organizations better triage and prioritize risks, we examined the frameworks and their questions/recommendations and extracted the risk factors each framework takes into account (see Table \ref{tab:industrialFrameworks} and Fig. \ref{fig:Riskfactors}).
Despite their respective focus (e.g., I14 focuses on hazards while touching vulnerability and exposure),
all 16 frameworks consider potential vulnerability, and 15 frameworks cover hazard and exposure.
However, mitigation risks are significantly underemphasized and only covered by AU NSW AI Assurance Framework (I7) and ECP's AI impact assessment framework (I15).
Even for these two frameworks that consider mitigation risks, they do not provide a comprehensive assessment rather briefly mention such risks. For example, in I7: ``\textit{Are there any residual risks?}", and in I15: ``\textit{Considering planned mitigations, could the AI system cause significant or irreversible harms?}".

\begin{center}
\begin{tcolorbox}[breakable, colback=gray!10,
colframe=black, 
width=\columnwidth,
arc = 1mm,
boxrule=0.5 pt,
]
\textbf{Finding 1 to RQ2.2}: The collected AI risk assessment frameworks generally do not clearly distinguish among different types of risk factors (e.g., hazard, exposure, vulnerability, and mitigation risk) to effectively identify and mitigate crucial AI risks. Although collected frameworks categorically encompass these factors to some degree, they may focus on particular factors while neglecting others. In particular, mitigation risks are often significantly neglected. This can result in potential failure to identify and mitigate crucial AI risks, which can have severe consequences for organizations and society at large.
\end{tcolorbox}
\end{center}

\begin{center}
\begin{tcolorbox}[breakable, colback=gray!10,
colframe=black, 
width=\columnwidth,
arc = 1mm,
boxrule=0.5 pt,
]
\textbf{Finding 2 to RQ2.2}: The collected AI risk assessment frameworks provide some information on assessment procedures but often lack clarity in specifying the inputs/outputs, stakeholders, and resources required at each step of the assessment process.
\end{tcolorbox}
\end{center}

\begin{center}
\begin{tcolorbox}[breakable, colback=gray!10,
colframe=black, 
width=\columnwidth,
arc = 1mm,
boxrule=0.5 pt,
]
\textbf{Finding 3 to RQ2.2}:
Many AI risk assessment frameworks list assessment measures (i.e., questions or checklists) without considering their interconnections or dependencies. As a result, the assessment process may be inefficient, which can hinder the identification and mitigation of AI risks.
\end{tcolorbox}
\end{center}

\subsubsection{RQ2.3: What are the outputs?}
This section discuss the outputs of the procedural frameworks.

Whether the output of a documented report is specified or not, the outputs of the procedural frameworks are, or at least should be, risk/impact assessment reports.
Some frameworks, such as I2, which creates a visualization of the risk level correlated to the RAI principles, and I3, which calculates risk and mitigation scores for various risk areas and generates the level of impact, generate reports automatically.
For I10, assessors must manually generate a report based on the questionnaire and their answers to the questions. The other procedural frameworks (I4, I5, I7-I9, I11, I13-I16) serve as (self-)assessment tools to guide assessors in identifying risks and do not require the preparation of a report.
However, since assessors should clearly document all the answers and the related questions when using the procedural frameworks, the processes result in documented assessment reports.


\begin{center}
\begin{tcolorbox}[breakable, colback=gray!10,
colframe=black, 
width=\columnwidth,
arc = 1mm,
boxrule=0.5 pt,
]
\textbf{Finding to RQ2}: Many collected AI risk assessment frameworks do not provide concrete mitigation solutions, or lack a structured way to present them, which can make it challenging for organizations to effectively address identified risks. As organizations increasingly rely on these frameworks for potential mitigation solutions, addressing this limitation is crucial for ensuring the effectiveness of AI risk assessments.
\end{tcolorbox}
\end{center}

%% file: 4Discussion.tex
\subsection{On C$^2$AIRA: Framework concreteness and connectedness}
\begin{figure*}[htbp]
\centering
\includegraphics[width=0.86\textwidth]{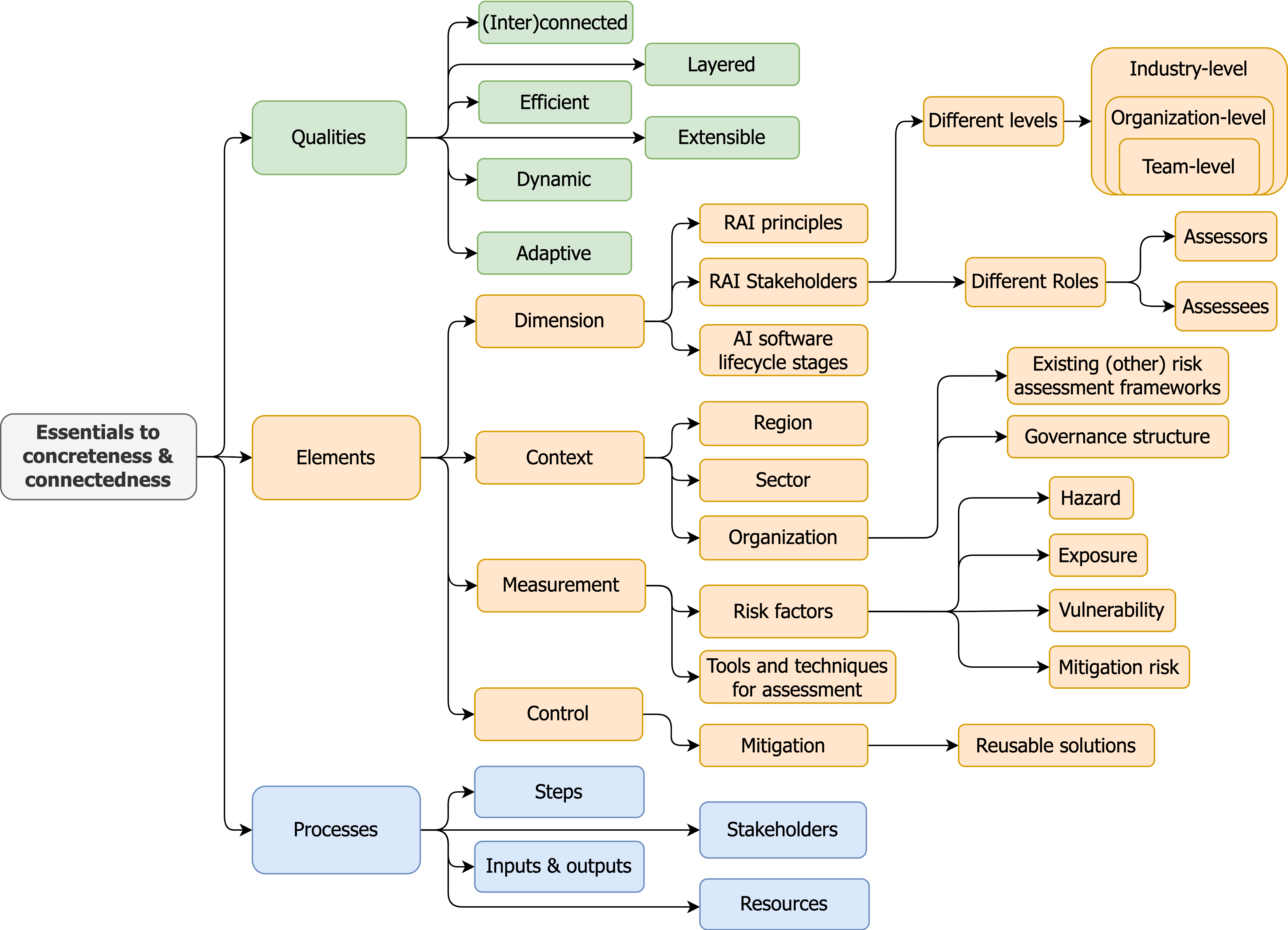}
\caption{Essentials to building a concrete and connected AI risk assessment (C$^2$AIRA) framework.} \label{fig:concreteness}
\end{figure*}

\subsubsection{Relative concreteness}
AI risk assessment frameworks may appear concrete at one level but too abstract for the next. For example, management teams may consider certain assessment questions concrete, while development teams may find them impractical. Additionally, even seemingly concrete checklists or templates for AI risk assessment may only be effective if assessors have a standardized and trustworthy approach to completing each item.
Therefore, it is crucial to have well-structured, concrete, and reusable solutions (e.g., design patterns \cite{lu2022responsible, lu2022responsible2}) in the lower level that align with higher-level practices such as governance guidelines to ensure a comprehensive and effective risk assessment.

\subsubsection{Trivialized concreteness}
Our mapping study reveals that many frameworks trivialize the concept of ``concreteness" by:
\begin{itemize}
    \item Applying existing assessment concepts to new AI-specific artifacts/processes without further specifying potential solutions. Examples include acknowledging the existence of AI risks and broadly mentioning that they need to be identified, documented, and mitigated.
    \item Identifying new concepts in AI and providing some sub-categorization without providing potential solutions. Examples include acknowledging bias as a common issue in AI systems and listing different sources of bias (e.g., data, algorithm), but not providing specified solutions to different biases.
    \item Identifying important AI risks and referring to potentially stale non-AI frameworks, which may not be suitable for addressing AI risks.
\end{itemize}

While higher-level frameworks may seem abstract, they do not necessarily trivialize concreteness. In fact, they can be helpful in pointing out areas where organizations can uplift their practices. However, it is important to ensure that these frameworks strike a balance between abstraction and concreteness via connectedness and clarity.
The key criteria for determining if a higher-level framework is concrete or not include:
\begin{itemize}
    \item Connectedness: whether high-level abstractions of potential assessment and/or mitigation measures are underpinned by lower-level concrete assessment techniques. This ensures that the framework is both comprehensive and grounded in practical considerations.
    \item Clarity: whether there is a clear understanding among higher-level stakeholders (e.g., management) about the inputs, processes, outputs, as well as required personnel and resources to complete the assessment. This understanding may not necessarily require technical expertise but rather an understanding of the trust placed in the lower-level concrete assessment techniques utilized.
\end{itemize}


\subsection{Essentials to ``concreteness" and ``connectedness"}

To achieve C$^2$AIRA, AI risk assessment frameworks need to have certain essential qualities, elements, and processes in place. Figure \ref{fig:concreteness} summarizes these requirements.

Firstly, a C$^2$AIRA framework must establish a clear connection between assessment and mitigation measures across different levels and stages of the AI development process. This connection should be reasonably underpinned, aligned, and connected, even if the measure itself is narrow-scoped and not directly covering different levels/stages. Secondly, the framework should be layered, organizing assessment and mitigation according to their dependencies on each other. This will result in a more efficient assessment process. Only the EU's ALTAI (I2) achieves a certain level of interconnectivity by providing an interactive online assessment tool where questions may vary depending on the previous answers. Thirdly, the framework should be extensible, dynamic, and adaptive, enabling it to be adapted and extended to more specific contexts. All of these qualities combine to enhance the efficiency of the assessment.

To be comprehensive, an AI risk assessment framework should cover different dimensions, contexts, measurements, and mitigations. Assessment and mitigation should be organized based on different RAI principles, RAI stakeholders, and AI-SDLC stages. The framework should also consider existing organizational governance structures and measures for integrating newly introduced AI governance with existing structures. For example, by establishing a dedicated RAI risk committee\cite{lu2022responsible} as part of the Board’s risk committee, it can help manage AI risks and promote a RAI culture within an organization.
Even if a framework focuses on a specific aspect (e.g., stage/principle-specific, designed for assessment instead of mitigation), it needs to be well connected with other frameworks that cover different aspects. Additionally, the framework should consider and specify contextual elements such as applicable regions, sectors, and compatibility with an organization's existing risk management processes and structures.
It is also important for the framework to consider different risk factors and present corresponding assessment and mitigation measures. Currently, mitigation risks are significantly neglected by existing frameworks. Reusable mitigation plans should be suggested in a structured way, along with their pros and cons (i.e., potentially mitigation risks) considered.

The framework should also specify the procedures required to conduct the AI risk assessment. This will help assessors and assessees from different levels better understand the inputs, processes, outputs, and required stakeholders and resources (e.g., data, tools, funds) for each step. However, only half (I1, I3, I4, I7, I10, I12, I14, I15) of the 16 frameworks provided such specifications to a certain extent. Additionally, seven out of the eight frameworks merely stated the steps needed to conduct the assessment, with only I4 specifying stakeholders involved in each step. None of the frameworks provides details on the resources required to complete each step.





\subsection{Threats to validity}
External threats: The term “AI risk assessment” and other similar terms such as “AI risk management” and “AI impact assessment” are often used interchangeably to refer to the identification, assessment, and mitigation of risks associated with AI. To address this issue, we conducted a thorough review of the literature to identify and include all relevant terms and keywords in our search strategy. However, we acknowledge that the search strategy may not have captured all relevant studies. Additionally, we only included publicly accessible AI risk assessment frameworks in our study, which may have excluded some frameworks used internally by organizations. Also, we acknowledge that our sample of 16 frameworks may not fully represent all existing AI risk assessment/management frameworks. Future studies could consider including additional academic literature and consulting with experts in the field to identify any additional frameworks that were not captured in our search.

Internal threats: To mitigate the risk of not finding all relevant studies, we conducted a rigorous search using defined keywords and supportive terms and performed snowballing to recover any missing studies. To address potential bias in data collection and synthesis, two researchers independently performed the tasks and reviewed and double-checked the results. Any inconsistencies were discussed and resolved through consensus.
Additionally, while our paper aims to include more``concrete" AI risk assessment frameworks, the definition of ``concreteness" can be subjective. To address this issue, we first conducted a small-scale pilot study to establish a preliminary level of agreement on the level of concreteness. During literature quality assessment, one author scored the collected literature based on their level of concreteness (1 for concrete, 0.5 for relatively concrete, and 0 for not concrete), which was then reviewed by two other authors. We acknowledge that this approach is not foolproof, and there may be alternative methods for evaluating concreteness.

%% file: 5RelatedWork.tex
Responsible AI has become an urgent topic of interest for both industry and academia, and managing AI risks is seen as a crucial means to achieve responsible AI. While numerous studies on responsible and ethical AI frameworks have been published, these frameworks tend to be more abstract than concrete in terms of risk assessment and management measures \cite{Floridi_Cowls_Beltrametti_Chatila_Chazerand_Dignum_Luetge_Madelin_Pagallo_Rossi_2018,Ong_2021}. Furthermore, while many studies on AI risks have been published in recent years, they often focus heavily on conceptualizing and categorizing AI risks, without providing specific solutions for risk assessment and mitigation \cite{perry2019ai, sotala2018disjunctive, ciupa2018conceptualizing, garvey2018ai}.

Recent years have seen an increasing number of proposals for more concrete and actionable solutions to managing AI risks. For instance, Zhang et al. proposed evaluating model risks by inspecting their behavior on counterfactuals \cite{10.1145/3477495.3531677}, while Schwee et al. introduced a toolchain for assessing privacy risks by taking in a model trained from the dataset to be shared and creating a privacy risk report \cite{schwee2020tool}. Yajima et al. showcased their work in progress on assessing machine learning security risks \cite{10.1145/3522664.3528613}. In addition, several studies have adopted or extended Failure Mode and Effect Analysis (FMEA) for assessing AI risks \cite{dominguez2021fails, rismani2021ai, li2022fmea}.

At the same time, there have been efforts to summarize related research on AI risk assessment and management. For example, the Institute for Ethical Machine Learning maintains a Github repository called ``Awesome AI Guidelines" where they summarize various responsible AI resources, including higher-level guidelines and frameworks, tools, standards, regulations, courses and more \cite{AIguidelines}. AlgorithmWatch also provides the AI Ethics Guidelines Global Inventory, which includes 167 RAI-related guidelines \cite{AIinventory} (as of March 2023). Notably, in January 2022, EY and Trilateral Research published a survey of AI risk assessment methodologies \cite{EYreport}. This survey presents a high-level overview of the global landscape of AI risk assessment, with the objective of providing RAI governors with noteworthy practices and regulations in the field. The report discusses: 1) regulations and legislation worldwide containing AI risk assessment related elements; 2) solutions to AI risk assessment by several international organizations; 3) standards related to AI risk management and governance; 4) a brief overview of part of the proposed approaches in industry and academia. While the report is categorically comprehensive, it lacks a detailed and systematic analysis of existing frameworks with more concrete assessment measures.

In contrast, the objective of this study is to provide RAI practitioners with a systematic summary of existing AI risk assessment frameworks and shed light on the future development of C$^2$AIRA frameworks.